\documentclass[11pt]{article}

\pdfoutput=1

\usepackage{cite}
\usepackage[dvipsnames]{xcolor}
\usepackage{amssymb,amsmath,amscd}
\usepackage{epsfig}
\usepackage{epstopdf}
\usepackage{latexsym}
\usepackage{graphicx}
\usepackage{booktabs}
\usepackage{bbm}
\usepackage[margin=15pt,small]{caption}
\usepackage{subcaption}
\usepackage{enumitem}
\usepackage{float}
\usepackage[toc]{appendix}
\usepackage[numbers,compress]{natbib}
\usepackage{tikz}
\usepackage[labelfont=bf]{caption}
\usepackage[export]{adjustbox}

\usepackage{braket,cellspace}
\usepackage{mathrsfs}
\usepackage{braket}
\usepackage{chngpage}
\usepackage{tabu}

\allowdisplaybreaks

\usetikzlibrary{positioning}
\definecolor{darkspringgreen}{rgb}{0.09, 0.45, 0.27}

\usepackage{datetime}
\usepackage[
      colorlinks=true,
      linkcolor=darkspringgreen,  
      urlcolor=darkspringgreen,    
      filecolor=darkspringgreen,     
      citecolor=darkspringgreen,
      linktocpage=true,
      pdfstartview=FitV,
      bookmarksopen=true     
      ]{hyperref}

\setlist{nolistsep}
\ifpdf
\fi

\let\oldbibliography\thebibliography
\renewcommand{\thebibliography}[1]{\oldbibliography{#1}
\setlength{\itemsep}{0pt}} 

\numberwithin{equation}{section} 

\newcommand{\mcite}[1]{\mbox{\cite{#1}}}

\begin{document}  

\begin{titlepage}

\begin{center} 

\vspace*{12mm}

{\LARGE \bf 
Solving Puzzles in Deformed JT Gravity:
\\[8pt]
Phase Transitions and Non-Perturbative Effects}

\bigskip
\bigskip
\bigskip
\bigskip

{\bf Clifford V. Johnson${}^{\dagger}$ and Felipe Rosso${}^{\dagger,\ast}$\\ }
\bigskip
\vskip 2mm
${}^{\dagger}$Department of Physics and Astronomy 
\\
University of Southern California \\
Los Angeles, CA 90089, USA  \\
\vskip 2mm
${}^{\ast}$Kavli Institute for Theoretical Physics \\
University of California\\
 Santa Barbara, CA 93106, USA \\
\bigskip
\tt{
johnson1@usc.edu,
felipero@usc.edu}  \\

\end{center}

\bigskip

\begin{abstract}
\noindent Recent work has shown that certain deformations of the scalar potential in Jackiw-Teitelboim gravity can be written as double-scaled matrix models. However, some of the deformations exhibit an apparent breakdown of unitarity in the form of a negative spectral density at disc order. We show here that the source of the problem is the presence of a multi-valued solution of the  leading order matrix model string equation. While for a class of deformations we fix the problem by identifying a first order phase transition, for others we show that the theory is both perturbatively and non-perturbatively inconsistent. Aspects of the phase structure of the deformations are mapped out, using  methods known to supply a non-perturbative definition of undeformed JT gravity. Some features are in qualitative agreement with a semi-classical analysis of the phase structure of two-dimensional black holes in these deformed theories.
\end{abstract}

\vfill

\end{titlepage}


\newpage

\setcounter{tocdepth}{2}
\tableofcontents

\section{Introduction}
\label{sec:introduction}

Low dimensional gravity theories have proven to be  extremely fruitful playgrounds for exploring important questions in quantum gravity that are substantially more difficult to address in higher dimensions. One of  the simplest examples is Jackiw-Teitelboim (JT) gravity \cite{Jackiw:1984je,Teitelboim:1983ux}, a two-dimensional  theory including a scalar (or ``dilaton'') with the following Euclidean action:
\begin{equation}\label{eq:39}
I_{\rm JT}[g_{\mu \nu},\phi]=
-S_0\,\chi 
-\frac{1}{2}\int d^2x\sqrt{g}\,
\phi(R+2)+\mbox{boundary terms} \ ,
\end{equation}
where $\chi$ is the Euler characteristic of the spacetime and $S_0$ a constant. This simple theory captures the low energy sector of both near-extremal black holes and branes \cite{Achucarro:1993fd,Fabbri:2000xh,Nayak:2018qej,Kolekar:2018sba,Ghosh:2019rcj,Sachdev:2019bjn} and the SYK model \cite{Sachdev:1992fk,Kit_SYK,Maldacena:2016hyu,Jensen:2016pah} (see ref.~\cite{Sarosi:2017ykf} for a review). Most importantly, the full quantum theory, where the spacetime is allowed to change topology, has been recently solved to all orders in perturbation theory in the topological expansion~\cite{Saad:2019lba}. This was accomplished by using a variety of powerful mathematical results \cite{Mirzakhani:2006fta,Eynard:2004mh,Eynard:2007kz} to show that the full topological expansion of the Euclidean partition function:
\begin{equation}
\label{eq:partfun}
Z(\beta)=\int_{-\infty}^{+\infty} \rho(E)e^{-\beta E} dE\ ,
\end{equation}
is reproduced by a double-scaled~\cite{Gross:1989aw,Douglas:1989ve,Brezin:1989ss} random Hermitian matrix model.

Given this success, it is reasonable to explore more complicated two-dimensional dilaton theories, obtained by deforming the JT gravity action (\ref{eq:39}). Studies of such deformations were recently initiated in refs.~\cite{Maxfield:2020ale,Witten:2020wvy}, by considering the following class of theories:
\begin{equation}\label{eq:49}
I_{\rm dJT}[g_{\mu \nu},\phi]=
I_{\rm JT}[g_{\mu \nu},\phi]-
\frac{1}{2}\int d^2x\sqrt{g}\,U(\phi)\ ,
\qquad {\rm where} \qquad
U(\phi)=2\sum_{i=1}^r\lambda_i
e^{-2\pi(1-\alpha_i)\phi}\ .
\end{equation}
For this to be a deformation of JT gravity,  $U(\phi)$ is  required to vanish at $\phi\rightarrow +\infty$ and so  $\alpha_i\le 1$. Further constraining $\alpha_i\in(0,1/2)$, the Euclidean partition function of this theory was computed in~refs.~\mcite{Maxfield:2020ale,Witten:2020wvy}, where the exponential terms in $U(\phi)$ were interpreted as inserting sharp defects in the ordinary JT path integral. In a similar fashion to the work of  ref.~\cite{Saad:2019lba}, the topological expansion of the partition function was shown to agree with that of a double scaled Hermitian matrix model.\footnote{The matching between the gravity and matrix model computations was made very explicit for arbitrary asymptotic boundaries and genus in ref.~\cite{Maxfield:2020ale}, using earlier results in refs.~\cite{Eynard:2007kz,Mertens:2020hbs}.} The matrix model expansion is defined from the leading genus spectral density~$\rho_0(E)$, obtained from the gravitational disc partition function after an inverse Laplace transform. All higher genus contributions are then obtained from~$\rho_0(E)$ {\it via}  recursion relations derived from the matrix model.

While for generic deformations $U(\phi)$ this procedure works nicely, it was noted in refs.~\cite{Maxfield:2020ale,Witten:2020wvy} that in certain cases the leading genus spectral density $\rho_0(E)$ provided by the disc partition function is not positive definite. Given that $\rho_0(E)$ is the seed of the recursion relation, starting from a sick spectral density results in an ill-defined perturbative expansion. This puzzle was left unresolved in refs.~\cite{Maxfield:2020ale,Witten:2020wvy}, but  it was speculated that either non-perturbative effects or phase transitions might resolve the issue. 

This paper was motivated by the desire to better understand these deformations and determine whether the problems can be resolved using recently developed techniques that provide an alternative formulation of the matrix model description of JT gravity in terms of certain combinations of minimal string models~\cite{Okuyama:2019xbv,Johnson:2019eik}. As emphasized in refs.~\cite{Johnson:2019eik,Johnson:2020heh,Johnson:2020exp}, this formulation is particularly well-suited to enable the extraction of non-perturbative physics. Fully specifying the matrix model is equivalent to determining a  function $u(x)$ for $x\in \mathbb{R}$, which is the potential of an associated quantum mechanics problem with $\hbar=e^{-S_0}$. While the function $u(x)$ satisfies a complicated non-linear ordinary differential equation (called the ``string equation''), its leading piece as $\hbar\rightarrow 0$, $u_0(x)$, satisfies a simple algebraic constraint. 

{\it A central result of our work is the observation that the non-positive spectral density~$\rho_0(E)$ arising in some JT gravity deformations is only a symptom of the sickness of the models, and the actual virus is a multi-valued potential~$u_0(x)$.} Sometimes the problem can be eradicated by identifying a phase transition, and at other times it cannot. In such cases, we give evidence that the non-perturbative physics (at least using the definitions employed here) does not exist. Furthermore, we explore the physics of the deformations in the semi-classical gravity approximation, the leading $\hbar$ behavior of the  matrix model, and the full non-perturbative definition, contrasting the three approaches and identifying several phenomena. We now summarize our core results in more detail, briefly explaining the  salient features.

\subsection{Summary of Results}

We start in \textbf{section \ref{sec:2}} by introducing the matrix model technology used throughout this work. An important issue we address is  the non-perturbative formulation. The Hermitian matrix model definition~\cite{Saad:2019lba} of JT gravity is afflicted by non-perturbative instabilities long known~\cite{Banks:1989df,Douglas:1990xv} to be present in certain double-scaled Hermitian matrix models. In ref.~\cite{Johnson:2019eik} an alternative matrix model definition of JT gravity was provided which is free from such instabilities and reproduces the perturbative physics of ref.~\cite{Saad:2019lba}. Here we present an extension of ref.~\cite{Johnson:2019eik}'s framework that is able to naturally incorporate the deformed JT gravity physics and serve as a non-perturbative definition. It reduces to the non-perturbatively stable definition when the deformations are removed and will therefore act as a sharp tool for determining if some deformations are non-perturbatively ill-defined (at least within a framework that works for ordinary JT gravity). The matrix model origins of the string equation are discussed.

\textbf{Section \ref{sec:3}} discusses a particular example of a deformation of JT gravity given by:
\begin{equation}
U(\phi)=2\lambda\left(
e^{-2\pi(1-\alpha_1)\phi}
-e^{-2\pi(1-\alpha_2)\phi}
\right)\ ,
\end{equation}
which satisfies $U(0)=0$. To start, we compute the Euclidean partition function in the semi-classical approximation and uncover an interesting variety of phase transitions of the two-dimensional black holes of the theory (see figures \ref{fig:5} and \ref{fig:6}). We continue by studying the full Euclidean partition function to leading order in genus (disc topology), computed in refs.~\cite{Maxfield:2020ale,Witten:2020wvy}. The leading genus spectral density obtained from the results in ref.~\cite{Witten:2020wvy} is given by:
\begin{equation}\label{eq:42}
\rho_0(E)=\frac{\sinh(2\pi\sqrt{E})}
{4\pi^2\hbar}+
\lambda
\frac{\cosh(2\pi\alpha_1\sqrt{E})-\cosh(2\pi\alpha_2\sqrt{E})}
{2\pi \hbar\sqrt{E}}\ ,
\end{equation}
where $\hbar\equiv e^{-S_0}$ and the threshold energy is $E_0=0$. Expanding this expression for small energies, one finds $\rho_0(E)\propto \sqrt{E}+\mathcal{O}(E^{3/2})$, where the proportionality factor in the leading term is negative for $\lambda>\lambda_c$, see (\ref{eq:54}).

To understand the origin of this issue from the matrix model perspective,  the model is reinterpreted as a particular combination~\cite{Okuyama:2019xbv} of minimal models, which enables a formulation to all orders in $\hbar$ and beyond \cite{Johnson:2019eik}. The central quantity in this formalism is the string equation, that determines a potential $u(x)$. The leading genus string equation associated to this model is given by:
\begin{equation}\label{eq:71}
\mathcal{R}_0[u_0,x]=
\frac{\sqrt{u_0}}{2\pi}
I_1(2\pi\sqrt{u_0})+
\lambda\big(
I_0(2\pi\alpha_1\sqrt{u_0})-
I_0(2\pi\alpha_2\sqrt{u_0})
\big)+x=0\ ,
\end{equation} (where $I_n$ is the $n$th modified Bessel function),
which provides an implicit definition of the leading genus potential $u_0(x)\equiv\lim_{\hbar \rightarrow 0}u(x)$. We first show that $\rho_0(E)$ in (\ref{eq:42}) goes negative whenever the corresponding potential $u_0(x)$ defined through equation~(\ref{eq:71}) becomes multi-valued, see figure~\ref{fig:1}. From this perspective, the issue is easily solved by picking the (unique) solution to the implicit equation~(\ref{eq:71}) that yields a single valued function~$u_0(x)$ in the region~${x<0}$. This modifies the leading  spectral density from~(\ref{eq:42}) to:
\begin{equation}\label{eq:44}
\rho_0(E)=
\frac{1}{2\pi \hbar}
\int_{E_0}^{E}
\frac{du_0}{\sqrt{E-u_0}}
(\partial_{u_0}\mathcal{R}_0)\ ,
\end{equation}
where $E_0\ge 0$ is determined from the largest solution to $\mathcal{R}_0[E_0,0]=0$. While for deformations for which $E_0=0$ the integral (\ref{eq:44}) reduces to (\ref{eq:42}), when $E_0>0$ the correct positive definite answer is instead obtained from the integral expression. The transition of $E_0$ between the two regimes is not analytic (see figure \ref{fig:3}) and corresponds to a phase transition which qualitatively matches with results obtained from the semi-classical analysis.\footnote{We should note that the semi-classical and matrix model computations apply to different $\alpha_i$ regimes, $\alpha_i\sim 1$ and $\alpha_i\in(0,1/2)$ respectively. While certain quantitative aspects do not match exactly (like the exact location and order of the transition), the qualitative agreement between the two approaches is quite remarkable.}

At this point we must comment on the resolution of this first puzzle and its relation to the calculations in ref.~\cite{Maxfield:2020ale}. In that work the disc partition function was shown to be given by (\ref{eq:44}) \textit{irrespective} of the value of $U(0)$. As a result, the gravitational computations in ref.~\cite{Maxfield:2020ale} already have the correct spectral density when $U(0)=0$, given by equation~(\ref{eq:44}) instead of equation~(\ref{eq:42}). However, the issue \textit{is} relevant from the perspective of ref.~\cite{Witten:2020wvy}, where the gravitational calculation with $U(0)=0$ unequivocally leads to the poorly defined result in (\ref{eq:42}). It would therefore be interesting to understand how the derivation of~(\ref{eq:42}) in ref.~\cite{Witten:2020wvy} must be modified in order to yield the correct result.

Section \ref{sec:3} concludes with a  computation of the full spectral density $\rho(E)$ including both  higher genus and non-perturbative contributions (see figures \ref{fig:15} and \ref{fig:14}). Non-perturbative effects generate corrections to $\rho_0(E)$ in (\ref{eq:44}) analogous to those previously computed for ordinary JT gravity in refs.~\cite{Johnson:2019eik,Johnson:2020exp}, with the difference that in this case $E_0$ can be non-zero.

\textbf{Section \ref{sec:4}}  repeats the analysis for an example of a  different  class of deformation of JT gravity. Here,  $U(0)\neq 0$, and:
\begin{equation}
U(\phi)=2\lambda e^{-2\pi(1-\alpha)\phi}\ .
\end{equation}
After computing the Euclidean partition function in the semi-classical approximation,  the full partition function to leading order in genus~\cite{Maxfield:2020ale,Witten:2020wvy} is studied. The spectral density in this case is given~by\footnote{We are using the same conventions as in ref.~\cite{Maxfield:2020ale}. This is obtained from ref.~\cite{Witten:2020wvy} by changing the integration variable in eq. (8.12) to $u_0=(1-s^2)E+s^2E_0$.} 
\begin{equation}\label{eq:45}
\rho_0(E)=
\frac{1}{2\pi \hbar}
\int_{E_0}^{E}
\frac{du_0}{\sqrt{E-u_0}}
(\partial_{u_0}\mathcal{R}_0)\ ,
\qquad
\mathcal{R}_0=
\frac{\sqrt{u_0}}{2\pi}I_1(2\pi\sqrt{u_0})+
\lambda I_0(2\pi\alpha\sqrt{u_0})+x\ ,
\end{equation}
with $E_0$ again determined from the largest solution to $\mathcal{R}_0[E_0,0]=0$. While the integral expression is the same answer as given in (\ref{eq:44}) for $U(0)=0$, when $U(0)\neq 0$ the result of the integral is not necessarily positive (see figure \ref{fig:16} for some examples). As in section~\ref{sec:3}, the issue originates in a multi-valued solution $u_0(x)$ to the leading genus string equation (\ref{eq:45}), but in this case in the region $x<0$ (see figure \ref{fig:16}). This makes the issue substantially different and ultimately results in a breakdown of the perturbative expansion of the matrix model. We also show that certain models which might appear to be well defined (since they have $\rho_0(E)$ positive definite) still suffer from this perturbative breakdown.

To determine whether the model can be made sense of even  when the perturbative expansion breaks down,  the system is studied non-perturbatively. We find that the non-perturbative definition also breaks down precisely when $u_0(x)$ becomes multi-valued in the region $x<0$. Hence, we argue that deformations of JT gravity for which $u_0(x)$ presents such multi-valuedness issues are both perturbatively and non-perturbatively unstable.

The paper concludes in \textbf{section \ref{sec:5}} with a brief discussion  and some thoughts about future directions. Several appendices contain details and results used in the main text.

\section{Matrix Models}
\label{sec:2}

This section  introduces some of  the random matrix model tools used throughout this work. The introduction will be rather brief (for longer treatments see {\it e.g.,} refs.~\cite{Ginsparg:1993is,DiFrancesco:1993cyw,Eynard:2015aea}) but will help establish some notation, and contextualize key aspects of the perturbative and non-perturbative formulations. 

\subsection{Double-Scaling Limit}
\label{sec:matrix-models-general}
Both   random Hermitian matrix models and  random complex matrix models will be relevant here, but it is traditional to begin with the former. 
The  model is defined through a probability measure $dH\,e^{-N\,{\rm Tr}\,V(H)}$ with $H$ an $N{\times}N$  Hermitian matrix and $V(H)$ a polynomial potential generalizing Wigner's prototype Gaussian case. Expectation values of matrix observables $\mathcal{O}$ are computed as:
\begin{equation}
\label{eq:observables}
\langle \mathcal{O} 
\rangle\equiv
\frac{1}{\mathcal{Z}}
\int dH\,\mathcal{O}\,e^{-\frac{N}{\gamma}\,{\rm Tr}\,V(H)}
\qquad {\rm where} \qquad
\mathcal{Z}\equiv 
\int dH\,e^{-\frac{N}{\gamma}\,{\rm Tr}\,V(H)}\ .
\end{equation}
One of the central observables of the theory is the spectral density
\begin{equation}
\label{eq:you-are-my-density}
\rho(E)\equiv 
\frac{1}{N}\Big\langle
\sum_{i=1}^N
\delta(E-\alpha_i)
\Big\rangle\ ,
\end{equation}
where $\alpha_i\in \mathbb{R}$ are the eigenvalues of $H$.\footnote{Denoted by $\alpha_i$ since the more standard notation for them, $\lambda_i$, is used for the deformation parameters of JT gravity.} In the $N\to\infty$ limit $\rho(E)$ becomes a smooth density function, beginning at some classical threshold energy $E_0$.

The central idea is to use random matrix models to compute universal properties of sums over random two-dimensional surfaces. Those surfaces can be seen~\cite{'tHooft:1973jz,Brezin:1978sv} to be  present in the definition above by expanding the integral $\mathcal{Z}$ in terms of Feynman diagrams (the quadratic term of $V(H)$ gives a propagator, higher order terms give vertices), which are graph duals of tesselations of the surfaces into polygons. The power $N^{\chi}$ is carried by a graph/polygonization with Euler characteristic~${\chi=2-2g-b}$ where~$g$ and~$b$ are the number of handles and boundaries, respectively. Universal physics from sum over surfaces ({\it i.e.,} properties that do not depend on model-dependent details such as whether triangles or pentagons were used for the discretization)  are picked out by taking the double-scaling limit~\cite{Gross:1989aw,Douglas:1989ve,Brezin:1989ss}.  This involves taking~${N\rightarrow \infty}$,  while simultaneously tuning the polynomial coefficients in the potential~$V(H)$ to a critical point dominated by smooth surfaces that are very large compared to the number of constituent polygons. The $1/N$ corrections  yield the topological expansion. The critical point is characterized by the rate of vanishing of the leading spectral density $\rho(E)$ at one of its endpoints (see {\it e.g.,} ref.~\cite{Dalley:1991zs}). The generic behavior is  Wigner's semi-circle law:~$\rho\sim (E-E_0)^\frac12$. An additional~$(k-1)$ zeros coincident at the endpoint~$E_0$ define the critical behavior corresponding to a distinct model~\cite{Kazakov:1989bc}, the~$k$th  model:~$\rho\sim(E-E_0)^{k-\frac{1}{2}}$.\footnote{These were termed ``multi-critical'' points in the old language, but we will avoid that usage here. In the string theory context, where the random surfaces represent string worldsheets, these models are the $(2k-1,2)$ ``minimal string theories", in the contemporary nomenclature.} 

Non-perturbative physics that lies beyond the topological $1/N$ expansion is accessible too.  
 A starting point for extracting it~\cite{Bessis:1980ss} is a family of polynomials $P_n(\alpha)=\alpha^n+\mbox{\rm lower powers}$ (where~${n=1\cdots N}$), that are orthogonal with respect to the measure~$d\mu(\alpha)={d\alpha\,e^{-\frac{N}{\gamma}\,V(\alpha)}}$:
\begin{equation}
\int d\mu(\alpha)P_n(\alpha)P_m(\alpha)\propto \delta_{mn}\ ,
\end{equation}
 and for which a recursion relation can be written  ({\it e.g.} for even $V(H)$) as: $\alpha P_n(\alpha)=P_{n+1}(\alpha) +R_nP_{n-1}(\alpha)$. All matrix model observables can be re-written in terms of the  recursion coefficients $R_n$ and identities satisfied by the defining matrix integral can be used to derive recursion relations for them. The $P_n(\alpha)$ actually define a finite dimensional Hilbert space description of the system, built from~${\ket{n}\propto P_n(\alpha)}$, with $\bra{n}m\rangle=\delta_{nm}$, within which observables, as discussed in equation~(\ref{eq:observables}), are associated with  operators.

In the double scaling limit, the Hilbert space supplied by the $P_n(\alpha)$ becomes infinite dimensional;~${\ket{n}\rightarrow \ket{x}}$ with $x\in \mathbb{R}$, and an effective quantum mechanical system emerges  from which matrix model observables can be computed~\cite{Banks:1989df,Gross:1989aw}. It is governed by a simple Schr\"odinger Hamiltonian:
\begin{equation}
\label{eq:Schrodinger}
\mathcal{H}[u]\equiv -\hbar^2\frac{\partial^2}{\partial x^2}+u(\hat{x})\ ,
\end{equation}
with  $\hat{x}\ket{x}=x\ket{x}$, and  $\hbar$ the scaling part of $1/N$ in the limit. The  potential $u(x)$ arises as the scaling part of the $R_n$ in the limit, and its functional form is determined by an ordinary differential equation called the ``string equation'' (to be discussed below) which is the scaling part of the aforementioned recursion relations for the $R_n$.  

An important observable is the expectation value of a ``macroscopic loop'' (a single boundary) of length $\beta$, which in this formalism is:
\begin{equation}\label{eq:68}
Z(\beta)\equiv\langle
{\rm Tr}\,e^{-\beta H}
\rangle=
\int_{-\infty}^{0}
dx
\braket{x|e^{-\beta \mathcal{H}[u]}|x}\ .
\end{equation}
It is important to notice the integration range for $x$. The double scaling limit captures the universal physics to be found in the infinitesimal neighbourhood of the spectral density's endpoint (where the critical behaviour was specified earlier). Recall that the index $n$ labeled the ordered $N$ energies, and so at large $N$ they are labeled by a continuous coordinate $X=n/N\in[0,1]$. The coordinate~${x\in(-\infty,+\infty)}$ describes the local physics of the $X=0$ endpoint, and in the present conventions they are related as  $X=0-x\delta^p$ (where $p$ is some positive power and $\delta\to0$ as $N\to\infty$). Crucially, the $x<0$ regime (integrated over in equation~(\ref{eq:68})) is the place where the surviving part of the spectral density classically has  support, and $x=0$ is  the location of the endpoint. Very importantly, note that {\it the full quantum mechanics is defined on the whole of the range of $x$}. The physics from the $x>0$ regime controls non--perturbative aspects of the system.

The part of the spectral density that survives in the double scaling limit, $\rho(E)$, can be written in this language too. Starting with equation (\ref{eq:68}) and inserting a complete set of eigenstates of the Hamiltonian, $\mathcal{H}\ket{\psi_E}=E\ket{\psi_E}$:
\begin{equation}\label{eq:69}
Z(\beta)= \int_{E_0}^{\infty} dE\, \rho(E) e^{-\beta E}\ ,
\qquad \qquad
\rho(E)= \int_{-\infty}^{0} dx\,|\psi_E(x)|^2\ ,
\end{equation}
where $E_0$ is the lowest  or ``threshold'' energy in the spectrum.

The macroscopic loop expectation value is denoted $Z(\beta)$ in the above (as done in equation~(\ref{eq:partfun})) because it is of the same structural form as JT gravity, {\it i.e.} a sum over all surfaces with a single boundary of fixed length~$\beta$.  Evidently, the central object in this formalism is the potential $u(x)$ that determines the Hamiltonian~(\ref{eq:Schrodinger}).  For the appropriate choice of $u(x)$, equation~(\ref{eq:69}) {\it will} be a definition of the JT gravity matrix model that is alternative to the recursion methods of ref.~\cite{Saad:2019lba}. The core idea is that the full non-perturbative behavior can be accessed from $u(x)$ obtained as the full solution to a differential equation, the string equation.

The all-orders perturbative expansion for $u(x)$ is also nicely packaged in the string equation. However, it is important to note here that the string equation arising from Hermitian matrix models is non-perturbatively unstable (this will be discussed in section~\ref{sec:2.1}). Therefore, the Hermitian matrix model string equation that we now write, should only be used perturbatively in $\hbar$ \cite{Gross:1989aw,Douglas:1989ve,Brezin:1989ss}:
\begin{equation}\label{eq:77}
\mathcal{R}[u,x]\equiv \sum_{k=0}^{\infty}
t_k\widetilde{R}_k[u]+x=0\ .
\end{equation}
The objects $\widetilde{R}_k[u]$ are  $k$th order polynomials in~$u(x)$ and its derivatives called the Gel'fand-Dikii polynomials~\cite{Gelfand:1975rn},  normalized so that the coefficient of~$u^k$ is unity. They satisfy a simple recursion relation~\cite{Gelfand:1975rn}:
\begin{equation}\label{eq:59}
\frac{2k+1}{2(k+1)}\widetilde{R}^{\prime}_{k+1}=
u\widetilde{R}^{\prime}_{k}
+\frac{1}{2}u'\widetilde{R}_{k}
-
\frac{\hbar^2}{4}\widetilde{R}^{\prime\prime\prime}_{k}
\ ,
\end{equation}
where $\widetilde{R}_0[u]=1$ and primes are derivatives with respect to $x$. 
The coefficients $t_k$  in equation~(\ref{eq:77}) indicate how much the $k$th model contributes to the form of  the  potential~$u(x)$ and essentially define the double scaled model. To construct the matrix model equivalent to JT gravity, a natural strategy is to determine the combination of $t_k$ that yields the leading perturbative $\rho_0(E)$,  known from disc level computations in JT gravity. This  determines the leading contribution to $u(x)$, denoted $u_0(x)$.  This is described next.

\subsection{Leading Perturbative Analysis}
\label{sec:2.1}

Writing a power series expansion for the potential $u(x)=\sum_{n=0}^{\infty}u_n(x)\hbar^n$ and inserting it into the string equation (\ref{eq:77}), a perturbative analysis of the physics can be performed. The leading  behavior (the classical limit in the quantum mechanics of equation~(\ref{eq:Schrodinger})) of the potential $u_0(x)$ satisfies the following simple algebraic equation:
\begin{equation}\label{eq:46}
\mathcal{R}_0[u_0,x]=
\sum_{k=0}^{\infty}t_k u_0^k+x=0\ .
\end{equation}
Subtleties regarding the implicit definition of $u_0=u_0(x)$ from this equation will play an important role in this paper. A simple and useful formula can be written for the  loop expectation value (the putative JT gravity partition function) $Z(\beta)$ defined in equation~(\ref{eq:68}) at leading (disc) order in genus ${Z_0(\beta)=\lim_{\hbar \rightarrow 0}Z(\beta)}$ in terms of $u_0(x)$. Using that the momentum~$\hat{p}=-i\hbar \partial_x$ and position operators~$\hat{x}$ commute to leading order in~$\hbar$ and inserting a complete set of momentum eigenstates~$\ket{p}$ we can solve the Gaussian integral in $p$ and find:
\begin{equation}\label{eq:56}
Z_0(\beta)=
\frac{1}{2\hbar \sqrt{\pi \beta}}
\int_{-\infty}^{0} dx\,
e^{-\beta u_0(x)}\ ,
\end{equation}
where we have used $\braket{p|x}=e^{ixp/\hbar}/\sqrt{2\pi \hbar}$. Changing the integration variable to $u_0$ and using (\ref{eq:46}), we can obtain the following formula for the spectral density $\rho_0(E)$ after applying an inverse Laplace transform:
\begin{equation}\label{eq:47}
\rho_0(E)=
\frac{1}{2\pi \hbar}
\int_{E_0}^{E}\frac{du_0}{\sqrt{E-u_0}}
(\partial_{u_0}\mathcal{R}_0)\ ,
\qquad {\rm where} \qquad
E_0=u_0(0)\ .
\end{equation}
This formula has played a central role in recent JT gravity explorations \cite{Okuyama:2019xbv,Johnson:2019eik}. In fact, in the mathematical literature this integral transform (when $E_0=0$) is known as the Abel transform, and in such cases,  it can be explicitly inverted. This is described in Appendix~\ref{sec:abel-inverse}, the final result is:
\begin{equation}\label{eq:48}
E_0=u(0)=0
\qquad \Longrightarrow \qquad
\mathcal{R}_0[u_0,x]=
2\hbar
\int_0^{u_0}
\frac{dE}{\sqrt{u_0-E}}
\rho_0(E)+x\ .
\end{equation}
This very useful simple formula allows us to write the leading genus string equation from the spectral density $\rho_0(E)$ and ultimately identify the coefficients $t_k$ that define the model. For example, in the case of undeformed JT gravity the result is~\cite{Okuyama:2019xbv,Johnson:2019eik}:
\begin{equation}
\label{eq:undeformed}
 \rho_0(E) = \frac{\sinh(2\pi\sqrt{E})}{4\pi^2\hbar} 
\qquad  \Longrightarrow \qquad  \mathcal{R}_0[u_0,x]=\frac{\sqrt{u_0}}{2\pi}I_1(2\pi\sqrt{u_0})+x =0\ , 
\end{equation}
and explicitly in terms of the $t_k$ this is:
\begin{equation}
\label{eq:teekay}
t_k=\frac{\pi^{2(k-1)}}{2k!(k-1)!}\ .
\end{equation}
As will become clear later, deformations of JT gravity will yield more general $\rho_0(E)$. This will imply different formulae for the $t_k$, and hence different matrix model definitions ({\it i.e.}, new combinations of the underlying minimal models). The resulting $u_0(x)$ can then be used as the seed for higher order corrections, as part of the  boundary conditions of the string equation that supplies non-perturbative physics. The latter will be described next.
 
\subsection{Non-Perturbative Completion}
\label{sec:non-perturbative}

A complete definition of JT gravity should  include a non-perturbative sector, and in this approach this means seeking contributions to $u(x)$  that cannot be captured in an expansion in $\hbar$. As already mentioned, the Hermitian matrix model string equation~(\ref{eq:77}) captures perturbation theory, but  fails to include consistent non-perturbative physics. This issue has a long history, and can be traced to the fact that the even $k$ minimal models defined by the Hermitian matrix model are individually non-perturbatively unstable \cite{Banks:1989df,Douglas:1990xv}. Since the Hermitian matrix model proposed by ref.~\cite{Saad:2019lba} includes (as reviewed in the previous section) the contribution of the $k$ even models, they inherit the instability.\footnote{A suggestion for how to avoid this issue was given in ref.~\cite{Johnson:2020exp}, obtained by thinking of the JT definition as a limit of a (stable) $k$ odd model, with the even models as perturbations of it. However, it is not clear if this will work. Moreover, the overall picture developed in this paper suggests that the needed solution to the differential equation simply does not exist. See the remark at the end of section~\ref{sub:4.3}.} 

Overall, this means that the non-perturbative definition of the double scaled Hermitian matrix model is unstable and must be modified. The function $u(x)$ must satisfy a different differential equation. In this work, the differential equation used can be obtained from two different perspectives: from an ensemble of \textit{complex} (instead of Hermitian) random matrices or from a scaling argument. Our proposal is a slight generalization of the non-perturbative definition of JT gravity recently given in ref.~\cite{Johnson:2019eik}, that incorporates cases  for which  $E_0\neq 0$. 

A different family of non-perturbately stable minimal models can be obtained from an ensemble of complex matrices $M$ with probability measure $dMdM^\dagger e^{-\frac{N}{\gamma}{\rm Tr}\,V(MM^\dagger)}$. The double scaling limit of this system was obtained in \cite{Morris:1991cq,Dalley:1992qg,Dalley:1992vr}, where much of the formalism described in section~\ref{sec:matrix-models-general} for Hermitian matrix models also applies (for a thorough exposition, see ref. \cite{Dalley:1992qg}). The crucial difference is given by the fact that the potential depends on the combination~$MM^\dagger\ge 0$, and so its eigenvalues~$\alpha_i$ are non-negative. The spectral density is therefore non-zero only on the positive real line. It is instructive to think of this as structurally the same problem provided by the Hermitian matrix model, but  with a ``wall'' at zero that prevents the energies from going below zero (presently we will extend the discussion to $E_0\neq0$). The procedure for tuning to a critical point is analogous to what was done before. In terms of the endpoint  of the distribution $\rho(E)$, for the $k$th model there are again $(k-1)$ additional zeros at the endpoint, but also the wall is present, which produces a key new feature. In terms of the variable $x$, the regime $x<0$ contains the physics of what happens all the way up to the wall, while the  $x>0$ sector yields the non-perturbative  (stabilising) physics supplied by the wall. It is  the presence of this latter regime that makes complex matrix models radically differ from  Hermitian matrix models, which contain no wall.\footnote{In fact, it is easy to see that if one insists on using  Hermitian matrix models, one can simply place a wall into such a model by hand, and reproduce the physics under discussion, as discussed in {\it e.g.}, ref.~\cite{Dalley:1991xx}. Alternatively since Hermitian matrix models are not really any better motivated for studying sums over two-dimensional surfaces than complex matrix models, one might as well start with complex matrix models which supply the wall ``naturally".}

\begin{figure}[t]
\centering
\begin{subfigure}{0.48\textwidth}
\includegraphics[width=\textwidth]{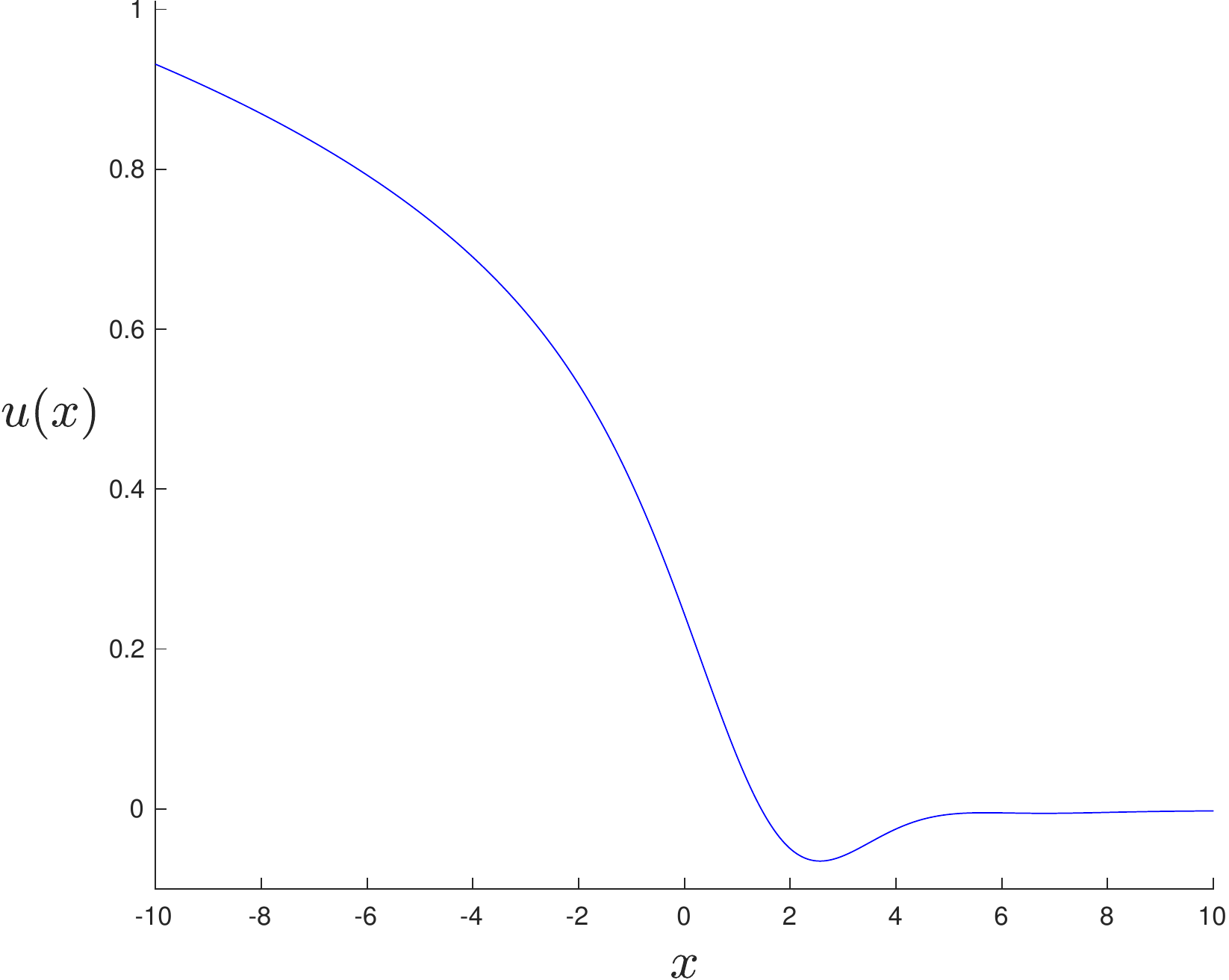}
\end{subfigure}
\begin{subfigure}{0.48\textwidth}
\includegraphics[width=\textwidth]{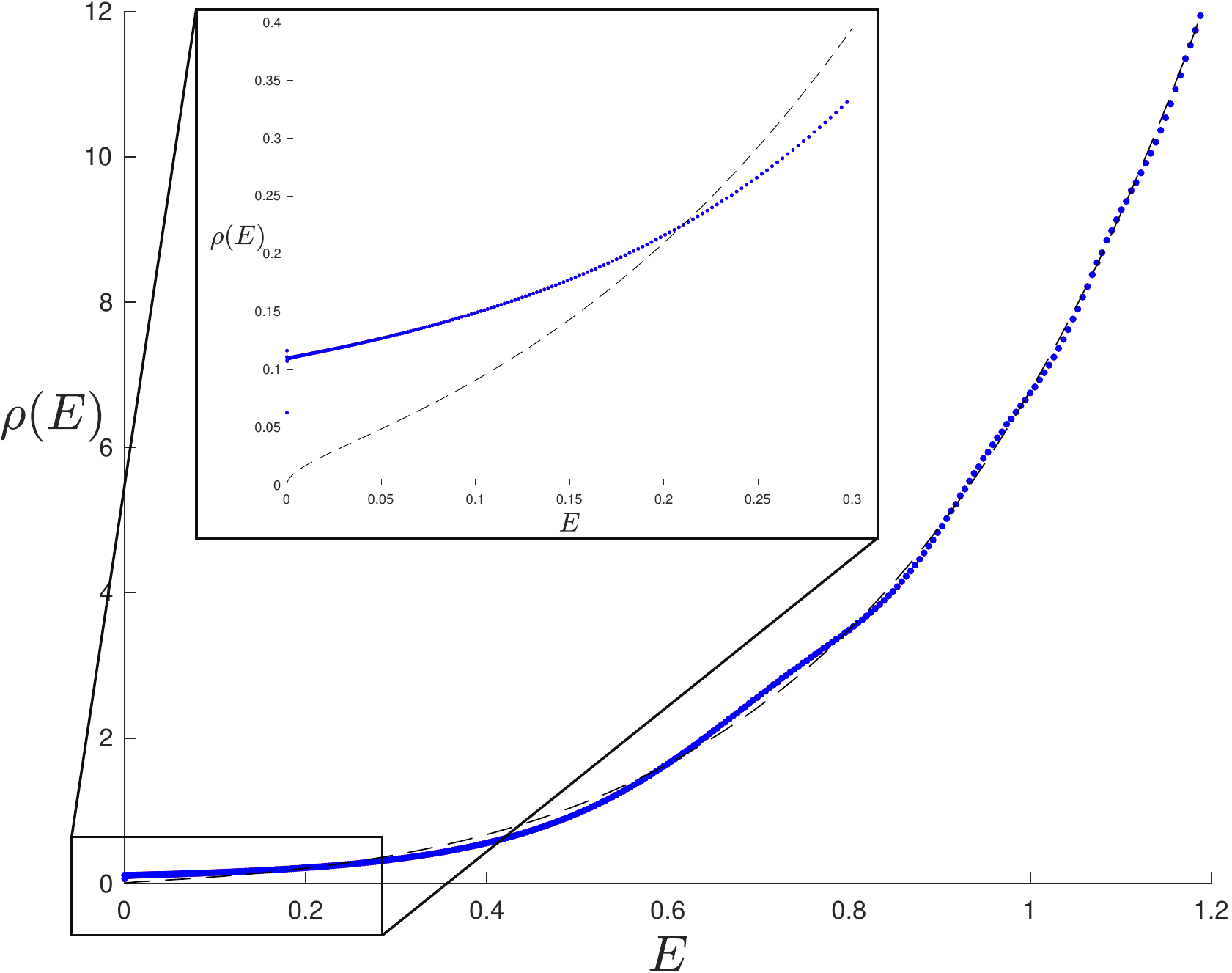}
\end{subfigure}
\caption{The non-perturbative potential (left) and spectral density $\rho(E)$ (right) for JT gravity computed in ref.~\cite{Johnson:2020exp} using equation~(\ref{eq:13}), truncated to $k=7$.  The dashed line is the disc level result for $\rho_0(E)$ given in equation~(\ref{eq:undeformed}). The inset shows the region near the origin, displaying a non-zero $\rho(E=0)$. Here, $\hbar$ is set to unity.}\label{fig:undeformed}
\end{figure}

The string equation that results from taking the double-scaling limit of these models is~\cite{Dalley:1991qg,Dalley:1992yi,Dalley:1991xx}:
\begin{equation}\label{eq:13}
(u-E_0)\mathcal{R}^2
-\frac{\hbar^2}{2}\mathcal{R}\mathcal{R}''
+\frac{\hbar^2}{4}(\mathcal{R}')^2
=0\ ,
\end{equation}
where $\mathcal{R}$ is defined as in (\ref{eq:77}). Note that the Hermitian matrix model string equation $\mathcal{R}=0$ is a solution of this equation, which results in the potential $u(x)$ agreeing to all orders in perturbation theory when expanding (\ref{eq:13}) in the $x<0$ regime, as is appropriate for JT gravity. The proposal of this paper is that for the deformations of JT gravity (as described in section~\ref{sec:introduction}) the equation~(\ref{eq:13}) gives the appropriate non-perturbative definition. For $E_0\neq 0$ it generalizes the previous proposal of ref.~\cite{Johnson:2019eik}, used in ref.~\cite{Johnson:2020exp} to explicitly compute the ordinary JT gravity non-perturbative potential $u(x)$ and spectral density~$\rho(E)$ shown in figure~\ref{fig:undeformed}.\footnote{Note that the individual $k$th models of  this subsection were (in minimal string language)  identified in ref.~\cite{Klebanov:2003wg} as the  $(4k,2)$ type 0A minimal strings. Refs.~\cite{Johnson:2020heh,Johnson:2020exp} combine them together in a {\it different} manner than done in  ref.~\cite{Johnson:2019eik} in order to non-perturbatively define various JT supergravity theories.}

Let us comment on the $E_0$ contribution in (\ref{eq:13}), that is the novel piece with respect to the proposal in ref. \cite{Johnson:2019eik}. The leading order behavior $u_0(x)$ obtained from (\ref{eq:13}) as $\hbar\rightarrow 0$ is given by
\begin{equation}
(u_0-E_0)\mathcal{R}_0^2[u_0,x]=0
\qquad \Longrightarrow \qquad
u_0(x)\equiv E_0
\qquad {\rm or} \qquad
\mathcal{R}_0[u_0,x]=0\ .
\end{equation}
If we want to have a single-valued and non-trivial function $u_0(x)$ that is defined in the whole real line~${x\in \mathbb{R}}$, none of the regimes is enough by itself, meaning there must be a transition between them. Since leading genus observables are determined by $u_0(x)$ in the region~$x<0$~(\ref{eq:56}), we should take~$u_0(x)$ defined from~$\mathcal{R}_0[u_0,x]=0$ when~$x<0$. In this way, leading genus observables computed from (\ref{eq:13}) agree with those obtained from the original Hermitian matrix model string equation. The full solution for~$u_0(x)$ is then given by 
\begin{equation}\label{eq:19}
u_0(x)=
\begin{cases}
\begin{aligned}
\,\,\,
\mathcal{R}_0[u_0&,x]=0
\quad \,\,\, \ , \qquad x<0\ , \\[4pt]
\,\, &\, E_0 
\qquad \quad \,\,  \ , \qquad x>0\ .
\end{aligned}
\end{cases}
\end{equation}
In order to have a continuous solution, we must fix~$E_0$ as~$\mathcal{R}_0[u_0(0)=E_0,x=0]=0$, which gives the following condition
\begin{equation}
\mathcal{R}_0[E_0,0]=\sum_{k=0}^{\infty}t_kE_0^k=0\ .
\end{equation}
The threshold value~$E_0$ is the energy at the endpoint of the spectral density, where the wall is, at~$x=0$. The origin of the shifted equation~(\ref{eq:13}), which first appeared in ref.~\cite{Dalley:1992yi}, can be understood as a simple modification of the complex matrix model by shifting the matrix combination~${MM^\dagger\rightarrow (MM^\dagger+E_0)}$, so that the eigenvalues lie above the threshold~$E_0$. The region~$x>0$ is {\it beyond} the wall, and so classically~$u_0(x)$ should stay frozen at~$E_0$, the value it reached at~$x=0$. The boundary conditions used to solve the full string equation (\ref{eq:13}) are obtained from the asymptotic expansions for~${x\rightarrow \pm \infty}$ starting from each of the two regimes in (\ref{eq:19}).

There is a different way of motivating the full differential equation (\ref{eq:13}), following the work in refs.~\mcite{Dalley:1991qg,Dalley:1991xx}. Instead of going through the double scaling limit of complex matrices, one seeks an equation that has the same perturbative physics as given by the Hermitian matrix model (already known to agree with JT gravity perturbatively) but then assumes that it preserves the following two established properties of $u(x)$ not just in perturbation theory but  {\it non-perturbatively}: 
\begin{equation}\label{eq:60}
\begin{aligned}
1.&\,{\rm KdV\,\,flow:}
\qquad \qquad \quad \,\,
(k+1)
\frac{\partial u}{\partial t_k}=
\frac{\partial \widetilde{R}_{k+1}[u]}
{\partial x}\ , \\[3pt]
2.&\,{\rm Scaling:}
\hspace{21mm}
u\big(sx;\lbrace s^{2k+1}t_k \rbrace,s^{-2}E_0\big)=
s^{-2}u(x;\lbrace t_k\rbrace,E_0)\ ,\\[4pt]
\end{aligned}
\end{equation}
where the first line refers to the explicit dependence of the potential $u(x)$ on the coefficients $\lbrace t_k \rbrace$. Its evolution  as a function of $t_k$ is governed by the KdV flow equation~\mcite{Douglas:1990dd,Banks:1989df}. The second line is a simple scaling relation that specifies how $u(x,\{t_k\},E_0)$ changes  with a rescaling of  the parameters by~$s$.\footnote{The simplest way to obtain this is by looking at the leading genus string equation $\mathcal{R}_0[u_0,x]=0$ in (\ref{eq:46}). By rescaling the parameters $(x,t_k,E_0)$ by $s$ as indicated, we obtain the scaling relation in (\ref{eq:60}).} Appendix \ref{sec:alt-derivation} reviews how to  use these conditions,  with~(\ref{eq:19}), 
to arrive at the string equation~(\ref{eq:13}).


All things considered, the non-perturbative definition given in this section is a well motivated proposal that allows for explicit computations of physical quantities in deformations of JT gravity.

\section{Deformed JT Gravity: Model A}
\label{sec:3}

In this section we study a particular deformation of JT gravity in (\ref{eq:49}), where the dilaton potential is deformed according to
\begin{equation}\label{eq:70}
U(\phi)=2\lambda\left(
e^{-2\pi(1-\alpha_1)\phi}
-e^{-2\pi(1-\alpha_2)\phi}
\right)\ ,
\end{equation}
where $U(0)=0$ and for simplicity we assume $\alpha_1<\alpha_2$. In the following sections we analyze the Euclidean partition function using the semi-classical approximation as well as the matrix model description. Several interesting characteristics of this theory will be  uncovered. Overall, it will emerge  as both perturbatively and non-perturbatively well defined.

\subsection{Semi-Classical Approximation}
\label{sub:3.1}

A standard semi-classical analysis of the deformed JT gravity theory involves computing the Euclidean partition function through a saddle point approximation:
\begin{equation}\label{eq:37}
Z(\beta)=
\int Dg\,D\phi\,e^{-I_{\rm dJT}[g_{\mu \nu},\phi]}
\simeq
\sum_{i}
e^{-I_{\rm dJT}[g_{\mu \nu}^{(i)},\phi^{(i)}]}
+\dots\ ,
\end{equation}
where $g_{\mu \nu}^{(i)}$ and $\phi^{(i)}$ are on-shell solutions to the equation of motion with renormalized boundary length $\beta$. This procedure yields a good approximation when $\alpha_i\sim 1$ (see appendix D of ref.~\cite{Maxfield:2020ale}). 

The first step in computing (\ref{eq:37}) is solving the equations of motion with a dilaton potential given by $W(\phi)= 2\phi+U(\phi)$ and the appropriate boundary conditions. Luckily, this has been analysed a long time ago and the general solution can be written as \cite{LouisMartinez:1993cc,Nappi:1992as,Witten:2020ert}:
\begin{equation}\label{eq:62}
ds^2=f(r)(dt/2)^2+\frac{dr^2}{f(r)}\ ,
\qquad \quad
f(r)=\int_{\phi_h}^r
d\phi\,W(\phi)\ ,
\qquad \quad
\phi(r)=r\ ,
\end{equation}
where $r\ge \phi_h$ and $t\sim t+\beta$.\footnote{The extra factor of two in the time coordinate is such that the boundary conditions are satisfied with $\beta$ matching with the conventions in ref.~\cite{Maxfield:2020ale}.} Black hole solutions are labeled by $\phi_h$, which must satisfy the following constraint in order to have a well defined signature:
\begin{equation}\label{eq:35}
\int_{\phi_h}^\phi
d\phi'\,W(\phi')\ge 0
\,\,\,\, {\rm for\,\, \,all\,\,\,}
\phi\ge \phi_h\ ,
\end{equation}
which in particular implies $W(\phi_h)\ge 0$. Evaluating the on-shell action (\ref{eq:37}), a standard computation yields the following thermodynamic quantities \cite{Witten:2020ert}:
\begin{equation}\label{eq:53}
T=\frac{W(\phi_h)}{2\pi}\ ,
\qquad \qquad
S=S_0+2\pi \phi_h\ ,
\qquad \qquad
E(\phi_h)=
\phi_h^2
-
\int_{\phi_h}^{\infty}d\phi \,U(\phi)\ .
\end{equation}
From this we can compute the specific heat and find the following stability condition for black hole solutions
\begin{equation}\label{eq:38}
\frac{dE}{dT}=4\pi
\frac{W(\phi_h)}{W'(\phi_h)}\ge 0
\qquad \Longrightarrow \qquad
{\rm stable\,\,black\,\,hole\,\,if\,\,}
W'(\phi_h)>0\ .
\end{equation}

\begin{figure}
\centering
\includegraphics[scale=0.32]{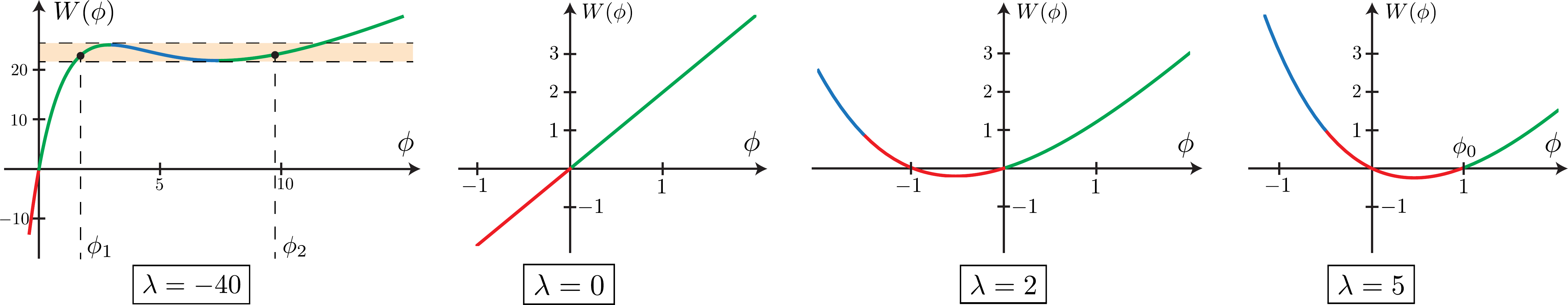}
\caption{Deformed JT gravity dilaton potential $W(\phi)=2\phi+U(\phi)$ in equation~(\ref{eq:70}) for the values $(\alpha_1,\alpha_2)=(0.9,0.95)$ and several values of $\lambda$. Segments of the curve in red, blue and green indicate whether the black hole solution with $\phi_h=\phi$ is nonexistent (violates (\ref{eq:35})), unstable (negative specific heat) or stable (positive specific heat).}\label{fig:5}
\end{figure}

Let us now specialize to the dilaton potential in equation~(\ref{eq:70}). The black hole solutions and their behavior can read off just by plotting the potential $W(\phi)=2\phi+U(\phi)$ for fixed $(\alpha_1,\alpha_2)\sim 1$ and~$\lambda$. Figure~\ref{fig:5}  shows sample characteristic behavior of the potential depending on the value of $\lambda$. Segments of the curve in red, blue and green indicate whether the black hole solution with $\phi_h=\phi$ is nonexistent (violates (\ref{eq:35})), unstable (negative specific heat) or stable (positive specific heat). The behavior can be quite different depending on the sign of $\lambda$, so each case is analysed separately.

\paragraph{Positive $\lambda$:} Since in this case the potential has $W(\pm \infty)=+\infty$ there are two branches of black hole solutions, one  stable and the other unstable. While at low enough temperatures there are only stable black holes, as we increase the temperature the unstable branch comes into play. In this regime, the favored black hole that dominates the partition function~(\ref{eq:53}) is the one with lower free energy. A simple analysis using (\ref{eq:53}) (see also section 3 of ref.~\cite{Witten:2020ert}) shows the stable black hole in the green branch always dominates the partition function over the unstable one in blue. As a result, the overall system is stable in the semi-classical approximation.

\begin{figure}
\centering
\includegraphics[scale=0.40]{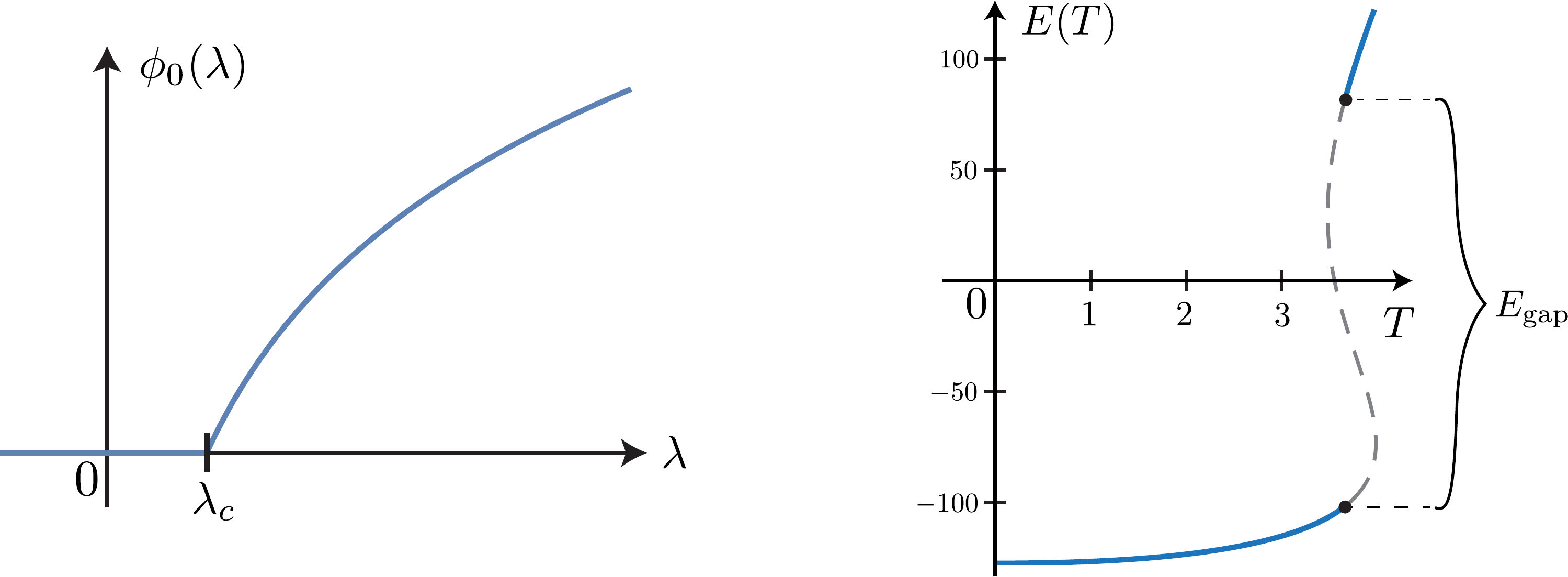}
\caption{The left diagram shows the  phase transition at~$\lambda=\lambda_c$ in equation~(\ref{eq:51}), where we plot~${\phi_h=\phi_0(\lambda)}$ obtained by numerically solving~$W(\phi_0)=0$. In the right diagram is plotted the black hole energy as a function of the temperature for~$\lambda=-40$, taking into account the phase transition (left diagram in figure \ref{fig:5}). We observe a gap in the spectrum generated by the phase transition at~$\lambda$ below~$\lambda_c'$ in (\ref{eq:72}).}\label{fig:6}
\end{figure}

Note that there is an interesting behavior for the zero temperature black hole solution $\phi_h=\phi_0$, determined by the largest root of~${W(\phi_0)=0}$. While for small enough $\lambda$ we always have $\phi_0=0$, there is a transition at some finite value $\lambda_c$ in which $\phi_0$ becomes non-zero and positive. The critical value~$\lambda_c$ can be easily computed as
\begin{equation}\label{eq:51}
W'(\phi)\big|_{\phi=0}=0
\qquad \Longrightarrow \qquad
\lambda_c=\frac{-1}{2\pi(\alpha_1-\alpha_2)}\ .
\end{equation}
Solving for the zero temperature black hole solution $\phi_0=\phi_0(\lambda)$ in a series expansion around $\lambda_c$ we find the following non-analytic behavior:
\begin{equation}\label{eq:52}
\phi_0(\lambda)=
\begin{cases}
\,\,\hspace{32mm} 0 \hspace{30mm}
\ , \qquad \lambda \le  \lambda_c\ ,\\[5pt]
\displaystyle
\,\,\frac{(1/\pi)}{(2-\alpha_1-\alpha_2)}
\left(\frac{\lambda-\lambda_c}
{\lambda_c}\right)+\mathcal{O}(\lambda-\lambda_c)^2
\ , \qquad \lambda \ge  \lambda_c\ .
\end{cases}
\end{equation}
The full function $\phi_0(\lambda)$ can be plotted numerically by solving $W(\phi_0)=0$, so that we get the left diagram in figure \ref{fig:6}. This corresponds to a phase transition of the zero temperature black hole solution. The order of the transition is obtained from the free energy $F=E-TS$, which at $T=0$ agrees with the energy (\ref{eq:53}):
\begin{equation}
F_0(\lambda)=E_0(\lambda)=
\phi_0(\lambda)^2-
\frac{\lambda}{\pi}
\left(
\frac{e^{-2\pi(1-\alpha_1)\phi_0(\lambda)}}
{1-\alpha_1}
-
\frac{e^{-2\pi(1-\alpha_2)\phi_0(\lambda)}}
{1-\alpha_2}
\right)\ ,
\end{equation}
with $\phi_0(\lambda)$ obtained from $W(\phi_0)=0$. From the series expansion (\ref{eq:52}) around $\lambda=\lambda_c$ we find that:\footnote{To get this expression we need to go to quadratic order in $\phi_0(\lambda)$. Solving for $\lambda=\lambda(\phi_0)$ from $W(\phi_0)=0$ we use that $\phi_0(\lambda)=(1/\lambda'(0))(\lambda-\lambda_c)-(\lambda''(0)/2\lambda'(0)^3)(\lambda-\lambda_c)+\cdots$ .}
\begin{equation}
E_0(\lambda)=
E_0(\lambda_c)
+\frac{(1/2\pi^2)}{(1-\alpha_1)(1-\alpha_2)}
\left(\frac{\lambda-\lambda_c}{\lambda_c}\right)
-
\begin{cases}
\hspace{26mm} 0 \hspace{26mm}
\ , \qquad \lambda \le \lambda_c\ , \\[5pt]
\displaystyle
\frac{(1/3\pi^2)}
{(2-\alpha_1-\alpha_2)^2}
\left(\frac{\lambda-\lambda_c}{\lambda_c}\right)^3+\dots
\ , \qquad \lambda \ge \lambda_c\ .
\end{cases}
\end{equation}
Since $E_0'''(\lambda_c^-)\neq E_0'''(\lambda_c^+)$, the semi-classical transition of the zero temperature black hole solution is third order.

\paragraph{Negative $\lambda$:} In this case  an interesting structure  arises for negative enough values of $\lambda$. The potential develops a local maximum and minimum, see leftmost diagram in figure \ref{fig:5}. This happens for $\lambda<\lambda_c'$ given by:
\begin{equation}\label{eq:72}
\begin{aligned}
\lambda_c'&=
\frac{1}{2\pi}\left[
(1-\alpha_1)
\left(
\frac{1-\alpha_1}{1-\alpha_2}
\right)^{\frac{2(1-\alpha_1)}{\alpha_1-\alpha_2}}
-(1-\alpha_2)
\left(
\frac{1-\alpha_1}{1-\alpha_2}
\right)^{\frac{2(1-\alpha_2)}{\alpha_1-\alpha_2}}
\right]^{-1}\ .
\end{aligned}
\end{equation}
It is straightforward to check that when $\lambda=\lambda_c'$ the potential has an inflection point
\begin{equation}
W'(\phi_c)\big|_{\lambda=\lambda_c'}=
W''(\phi_c)\big|_{\lambda=\lambda_c'}=0\ ,
\qquad \,\, {\rm where} \,\, \qquad
\phi_c=\frac{-1}{\pi(\alpha_1-\alpha_2)}
\ln\left[
\frac{1-\alpha_1}{1-\alpha_2}
\right]\ ,
\end{equation}
meaning that for $\lambda<\lambda_c'$ a local maximum and minimum develops. 

The behavior of the dilation potential $W(\phi)$ in this regime induces a phase transition of a different kind from the one discussed for positive $\lambda$. In the canonical ensemble (at fixed temperature~${T=W(\phi_h)/2\pi}$), we identify a shaded region in the left diagram in figure \ref{fig:5} where there is a finite range of temperatures with three black hole solutions. The dominant black hole in this regime is the one with the lowest free energy $F=E-TS$. From the equations in~(\ref{eq:53}), a simple analysis shows that seeking the dominant black hole produces  a discontinuous jump from $\phi_1$ to $\phi_2$, as indicated in the leftmost component of figure \ref{fig:5}. In the right diagram of figure \ref{fig:6} is plotted the energy as a function of the temperature (see eqs.~(\ref{eq:53})), taking into account the leap in $\phi_h$. At the critical temperature we observe a discontinuous jump in the energy. This generates a gap $E_{\rm gap}$ for which there are no dominant black hole solutions in a finite range of energies. This kind of first order phase transition (where three black holes compete) was first observed for charged AdS black holes in ref.~\cite{Chamblin:1999tk}. Along with other transitions, they have been recently explored in the current JT gravity context in ref.~\cite{Witten:2020ert}.

\subsection{Perturbative Genus Expansion}
\label{sub:3.2}

While the semi-classical analysis of the previous section showed a rich and interesting structure, it was obtained from the crude saddle point approximation to the Euclidean path integral (\ref{eq:37}). To compute $Z(\beta)$ more carefully we use the results in refs.~\cite{Maxfield:2020ale,Witten:2020wvy}, where it was shown the full partition function can be computed from a double scaled matrix model. In this subsection we focus on the leading genus contribution, $Z_0(\beta)$, given by the disc topology:
\begin{equation}\label{eq:73}
Z(\beta)=
\int Dg\,D\phi\,e^{-I_{\rm dJT}[g_{\mu \nu},\phi]}=\int_{\rm Disc} Dg\,D\phi\,e^{-I_{\rm dJT}[g_{\mu \nu},\phi]}
+\cdots = Z_0(\beta) +\cdots\ ,
\end{equation}
where we are neglecting additional higher genus and non-perturbative contributions that will be studied in the next subsection. The leading genus disc spectral density (\ref{eq:42}) derived in ref.~\cite{Witten:2020wvy}, that we rewrite here for convenience, is given by:
\begin{equation}\label{eq:41}
\rho_0(E)=\frac{\sinh(2\pi\sqrt{E})}
{4\pi^2\hbar}+
\lambda
\frac{\cosh(2\pi\alpha_1\sqrt{E})-\cosh(2\pi\alpha_2\sqrt{E})}
{2\pi \hbar\sqrt{E}}\ .
\end{equation}
This formula has a problem, since $\rho_0(E)$ is not always positive definite. This can seen  by expanding for small energies, where we find:\footnote{We have suggestively defined $\lambda_c$, which is different from the semi-classical definition in (\ref{eq:51}). We comment on the relation between the two at the end of this subsection.}
\begin{equation}\label{eq:54}
\rho_0(E)=
\left(
\frac{\lambda_c-\lambda}{2\pi\hbar \lambda_c}
\right)
\sqrt{E}+\mathcal{O}(E^{3/2})\ ,\qquad {\rm where} \qquad
\lambda_c=\frac{-1}{2\pi^2(\alpha_1^2-\alpha_2^2)}\ .
\end{equation}
Here $\lambda_c>0$ since we are assuming $\alpha_1<\alpha_2$. The spectral density goes negative for $\lambda>\lambda_c$ and small enough energies.\footnote{As already explained in the introduction, this issue does not arise from the perspective of ref.~\cite{Maxfield:2020ale}.} 

This certainly does not fit within the matrix model analysis and framework outlined in section~\ref{sec:2.1}. In particular, the definition of $\rho(E)$ in equation~(\ref{eq:you-are-my-density}) is manifestly positive, and large~$N$ and double-scaling limits cannot change that, as is clear from equation~(\ref{eq:69}).  To understand the origin of the issue from the matrix model perspective, consider the behavior of $u_0(x)$, defined from the leading genus string equation. Using equation~(\ref{eq:48}) we can easily compute $\mathcal{R}_0$ and find
\begin{equation}\label{eq:25}
\mathcal{R}_0[u_0,x]=
\frac{\sqrt{u_0}}{2\pi}
I_1(2\pi\sqrt{u_0})+
\lambda\big(
I_0(2\pi\alpha_1\sqrt{u_0})-
I_0(2\pi\alpha_2\sqrt{u_0})
\big)+x=0\ .
\end{equation}
To have a well defined single valued function~$u_0(x)$ on the real line~$x\in \mathbb{R}$, the implicit function theorem imposes the following constraint
\begin{equation}
u_0=u_0(x)\,\,{\rm for}\,\,x\in \mathbb{R}
\qquad \Longrightarrow \qquad
\partial_{u_0}\mathcal{R}_0\neq 0\ .
\end{equation}
From (\ref{eq:25}) we can easily check this condition is never satisfied, not even for undeformed  JT gravity ({\it i.e.,} $\lambda=0$). This might seem strange, since ordinary JT gravity has a  well defined disc spectral density (\ref{eq:undeformed}). However it is important to note that the leading genus partition function~$Z_0(\beta)$~(\ref{eq:56}) is built from the regime $x<0$. {\it Therefore we only need $\mathcal{R}_0[u_0,x]=0$ to define a unique single-valued function $u_0(x)$ in the range $x<0$.} 

\begin{figure}
\centering
\includegraphics[scale=0.33]{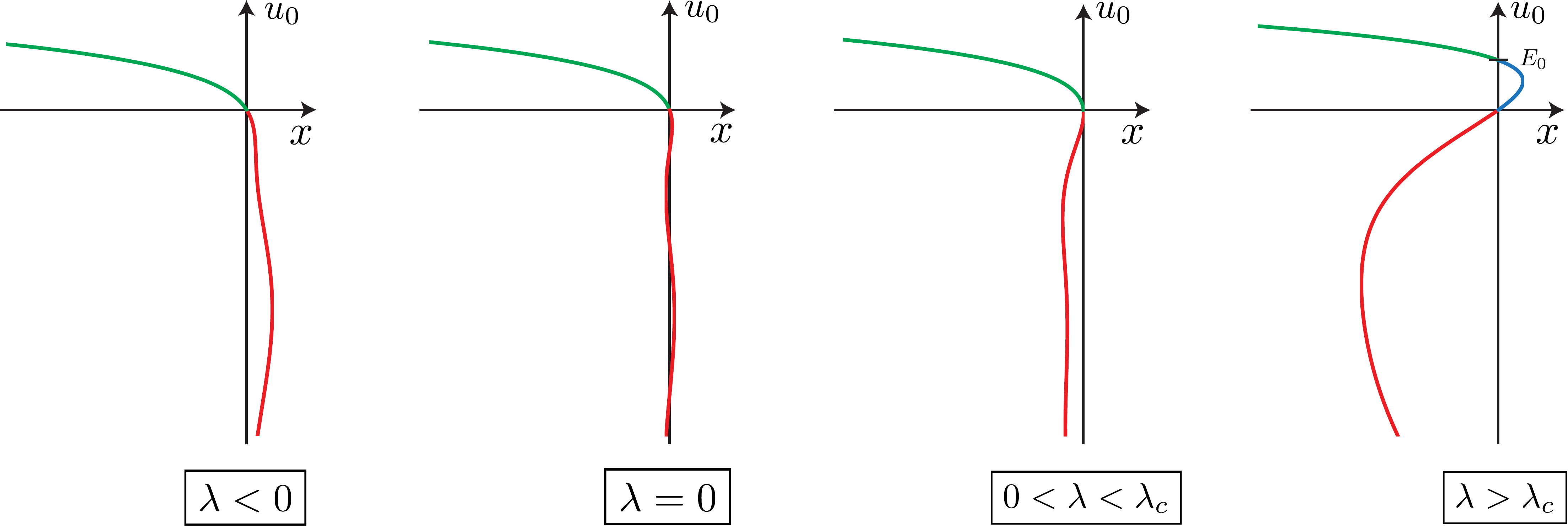}
\caption{Examples of the curve $x=x(u_0)$ obtained from (\ref{eq:25}) in the $(x,u_0)$ plane for several values of $\lambda$ and $(\alpha_1,\alpha_2)=(1/4,1/3)$. The green curve is the unique choice that yields a single valued function defined in the range $x<0$. While for $\lambda\le \lambda_c$ the green curve always satisfy $u(0)=0$, for $\lambda>\lambda_c$ we observe $u(0)=E_0>0$.}\label{fig:1}
\end{figure}

To understand the situation more clearly it is  useful to plot the curve $x=x(u_0)$ defined through equation~(\ref{eq:25}). The undeformed JT gravity case is the curve shown in the $\lambda=0$ diagram in figure~\ref{fig:1}. While for $u_0<0$ there are an infinite number of oscillations around $x=0$ (marked in red), for $u_0>0$ there is a single curve marked in green going from $u_0=0$ to $u_0=+\infty$. This green curve defines the unique single valued function $u_0(x)$ in $x<0$. As  can be seen from the other plots in figure \ref{fig:1}, the behavior for $\lambda<\lambda_c$ is quite similar: the upper (green) branch can be used to define a unique single-valued function defined in the range $x<0$. Using that in all these cases $u_0(0)=E_0=0$, the integral~(\ref{eq:47}) can be solved explicitly, recovering the starting result~(\ref{eq:41}), which is positive for such $\lambda$.

\begin{figure}
\centering
\includegraphics[scale=0.40]{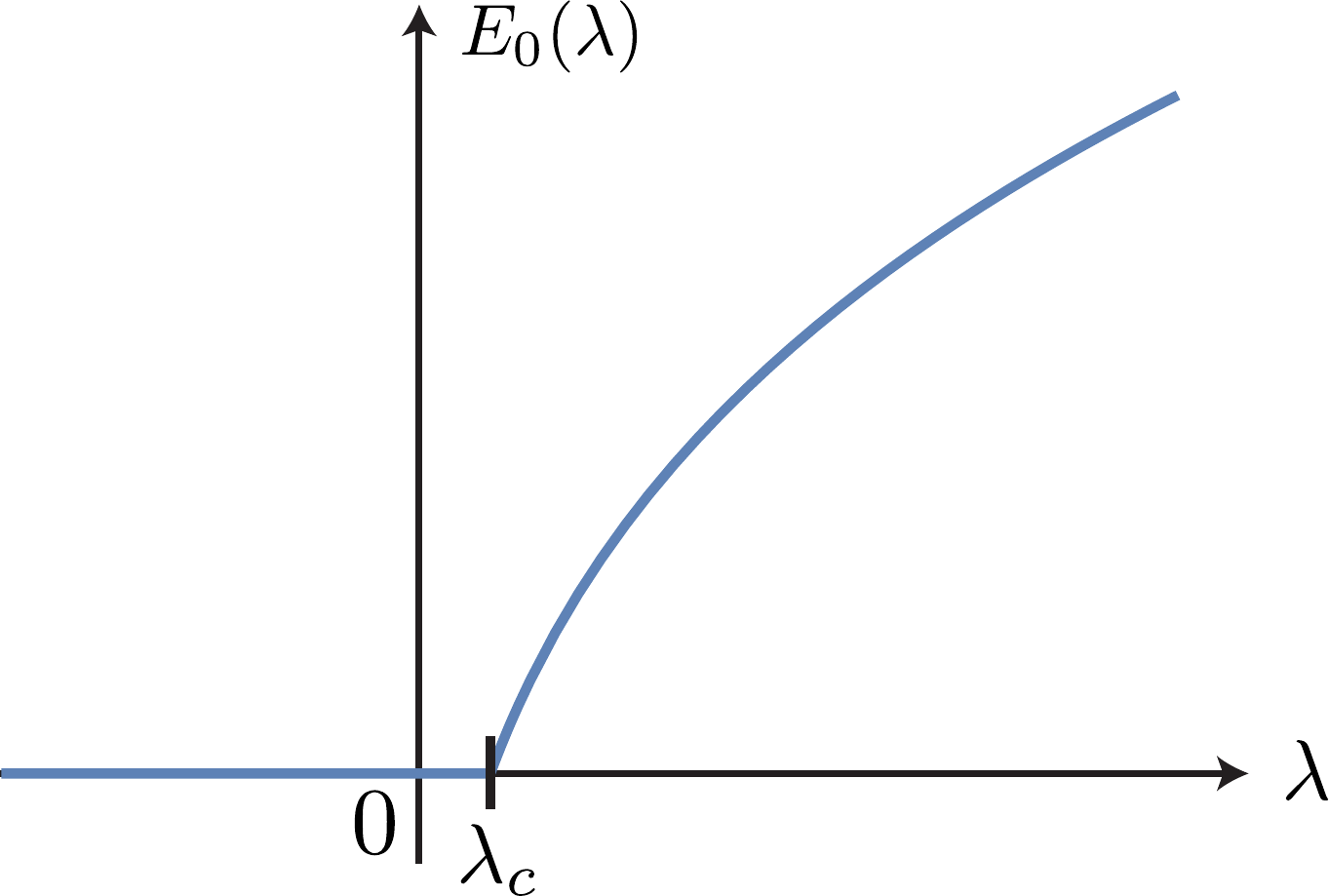}
\caption{Threshold energy $E_0$ as a function of $\lambda$ for $(\alpha_1,\alpha_2)=(1/4,1/3)$, obtained by numerically solving (\ref{eq:30}). For $\lambda=\lambda_c$ in (\ref{eq:54}) we observe a first order phase transition.}\label{fig:3}
\end{figure}

Consider now $\lambda>\lambda_c$. From the rightmost diagram in figure~\ref{fig:1} it is clear that there is an interesting feature: the value of $u_0(0)$ is no longer zero. The shift is caused by the fact that exactly at $\lambda=\lambda_c$ defined in (\ref{eq:54}), the derivative of the leading genus string equation vanishes at~$u_0=0$
\begin{equation}\label{eq:29}
(\partial_{u_0}\mathcal{R}_0)
\big|_{(u_0,\lambda)=(0,\lambda_c)}=0\ .
\end{equation}
Consider the curve starting at $u_0(0)=0$ in the rightmost diagram in figure~\ref{fig:1}, defining the multi-valued function $u_0(x)$ ({\it i.e.,}  including the blue section of the curve). Using it to compute $\rho_0(E)$ from the integral formula~(\ref{eq:47})  yields the non-positive result in equation~(\ref{eq:41}). On the other hand, if we proceed and define~$u_0(x)$ by picking the unique curve that results in a single valued function~$u_0(x)$ for~$x<0$, {\it i.e.} the green portion of the  diagram, the $\rho_0(E)$ obtained from integral~(\ref{eq:47}) is explicitly positive (since~${\partial_{u_0}\mathcal{R}_0\ge 0}$ in this case). The primary new feature is the value of $u_0(x)$ at the origin,  the threshold energy ${E_0=u_0(0)}$. For $\lambda>\lambda_c$, instead of having $E_0=0$ it is a non-zero value obtained from the largest solution to the leading genus equation at $x=0$
\begin{equation}\label{eq:30}
\mathcal{R}_0[u_0(0)=E_0,x=0]=
\frac{\sqrt{E_0}}{2\pi}
I_1(2\pi\sqrt{E_0})+
\lambda\big(
I_0(2\pi\alpha_1\sqrt{E_0})-
I_0(2\pi\alpha_2\sqrt{E_0})
\big)=0\ .
\end{equation}
This equation can be easily solved numerically to yield figure \ref{fig:3}. The behavior of $E_0$ as a function of $\lambda$ is non-analytic at $\lambda_c$ and corresponds to a phase transition. We can determine the order of the transition using (\ref{eq:30}) to expand $E_0(\lambda)$ on either side of $\lambda_c$, so that:
\begin{equation}
\label{eq:transition_A}
E_0(\lambda)=
\begin{cases}
\hspace{32mm} 0 \hspace{32mm}\,
\ , \qquad \lambda \le \lambda_c\ , \\[5pt]
\displaystyle 
\,\,
\frac{(2/\pi)^2}{(2-\alpha_1^2-\alpha_2^2)}
\left(\frac{\lambda-\lambda_c}{\lambda_c}\right)+\mathcal{O}(\lambda-\lambda_c)^2\,\,
\ , \qquad \lambda \ge  \lambda_c\ .
\end{cases}
\end{equation}
Since $E_0'(\lambda_c^+)\neq E_0'(\lambda_c^-)$  this is a first order phase transition. This phase transition solves the negativity of the leading genus spectral density, which for $\lambda>\lambda_c$ is given by (\ref{eq:47}) with $u(0)=E_0\neq 0$ instead of the problematic formula~(\ref{eq:41}).

It is interesting to  compare these leading order matrix model results with those obtained in the previous subsection from the semi-classical gravity analysis. The most important similarity is given by the fact that in both cases we observe a phase transition of the threshold energy~$E_0$ for some positive value of $\lambda$. By comparing figures \ref{fig:5} and \ref{fig:1} we also note some resemblance in the mechanisms triggering the transition. There are however important quantitative differences, since the exact location of the transition $\lambda_c$ in (\ref{eq:51}) and (\ref{eq:54}) does not agree. Moreover, while the matrix model transition is first order, the semi-classical analysis yields a third order transition. When doing this comparison,  keep in mind there is no overlap between the $\alpha_i$ regimes of validity of each computation. Moreover, the saddle-point approximation of the path integral~(\ref{eq:37}) does not take into account quantum corrections that might be important at low energies. All things considered, it is quite pleasing (and perhaps surprising) that the semi-classical analysis captures the transition of the vacuum solution at all. 

On the other hand, the semi-classical gravity analysis yielded a phase transition among black hole solutions that is present at negative~$\lambda$ (below~$\lambda_c^\prime$ in~(\ref{eq:72})). It suggested the existence of a gap~$E_{\rm gap}$ opening up in the available energy of black holes. There is no sign of such a gap in the spectrum in the matrix model analysis at disc level, but what about higher orders in $\hbar$ in the perturbative expansion?

A simple way of showing perturbative corrections are not able to account for $E_{\rm gap}$ is by computing the first genus correction to the disc spectral density $\rho_0(E)$ in (\ref{eq:41}). The partition function at this order is given in equation (9.19) of ref.~\cite{Witten:2020wvy}. Carefully applying an inverse Laplace transform (dealing with divergences in the same way as in footnote 9 of ref.~\cite{Maxfield:2020ale}) we find the corrections to spectral density are of the form $\hbar c_0/E^{3/2}+\hbar c_1/E^{5/2}$ for some constants $c_0$ and $c_1$. Contributions of this form are not able to explain the energy gap $E_{\rm gap}$, which should arise from a low energy behavior given by:
\begin{equation}
\rho(E)\sim
\sqrt{E-E_{\rm gap}}
=\sqrt{E}
-\frac{E_{\rm gap}}{2\sqrt{E}}
-\frac{E_{\rm gap}^2}{8E^{3/2}}
+\mathcal{O}(1/E^{5/2})\ .
\end{equation}
Neither the leading genus spectral density $\rho_0(E)$ in (\ref{eq:41}) nor its first correction contribute with the term $\sim 1/\sqrt{E}$ that is necessary to resum the series and obtain the gap. Overall, the energy gap~$E_{\rm gap}$ identified in the semi-classical analysis is not observed when computing the Euclidean partition function including contributions of arbitrary genus.

\subsection{Non-Perturbative Effects}
\label{sub:3.3}

So far, the Euclidean partition function of the theory has been studied in one or other approximation scheme: either the semi-classical gravity limit~(\ref{eq:37}) or the perturbative expansion of the matrix model~(\ref{eq:73}). Using our  double-scaled matrix model definition presented in section~\ref{sec:non-perturbative}, we now compute $Z(\beta)$ \textit{exactly}, including all higher order genus and non-perturbative effects. 

The first step is to identify the coefficients $t_k(\lambda)$ in (\ref{eq:77}) that define the matrix model. These can be obtained by expanding the leading genus string equation~(\ref{eq:25}) in a power series in $u_0$, giving:
\begin{equation}
\mathcal{R}_0[u_0,x]=\sum_{k=1}^{\infty}t_k(\lambda)u_0^k+x\ ,
\qquad {\rm  where} \qquad
t_k(\lambda)=\frac{\pi^{2(k-1)}}{2k!^2}
\big(
k+2\pi^2\lambda(\alpha_1^{2k}-\alpha_2^{2k})
\big)\ .
\end{equation}
Using this, the combination $\mathcal{R}[u,x]$ in (\ref{eq:77}) can be  constructed, and hence the non-linear ordinary differential equation (\ref{eq:13}) that determines the potential $u(x)$.  The boundary condition is the full~${\hbar=0}$ solution for $u_0(x)$, which is given by the $u_0(x)$ discussed in the previous section for $x<0$, and for~${x\geq0}$ by~$u_0(x)=E_0$.  Since all the $t_k(\lambda)$ are non-zero, the differential equation is of infinite order, but (as demonstrated in the case of $\lambda=0$ in ref.~\cite{Johnson:2020exp}) there is a meaningful sense in which it can be truncated to finite order, by including all $t_k$ contributions up to a maximum, $t_{k_{\rm max}}$:
\begin{equation}\label{eq:21}
\mathcal{R}=\sum_{k=1}^{\infty}t_k(\lambda)\widetilde{R}_k[u]+x
\qquad \xrightarrow[{\rm truncation}]{} \qquad
\mathcal{R}^{(k_{\rm max})}=
\sum_{k=1}^{k_{\rm max}}
t_k(\lambda)\widetilde{R}_k[u]+x\ .
\end{equation}
This is a controlled truncation, as can be seen from the fact that the $t_k$ rapidly decrease in relative size as $k$ increases.  The truncation gives a good estimate for $E_0$, $\lambda_c$, and moreover gives an energy spectrum that matches well to the exact one. Adding higher $k$ improves the truncation accuracy. It was found that $k_{\rm max}=7$ gave good accuracy for the range of energies where non-perturbative effects are important. This required a solution of  a highly non-linear 14th order differential equation, which turns out to be possible.\footnote{In practice, a first derivative of the string equation (and then dividing by ${\cal R}$) results in an equation that is structurally less complicated, so a 15th order equation was solved for $k_{\rm max}=7$. The boundary conditions, obtained by fixing the appropriate number of derivatives $\partial_x^{n} u(x)\equiv u^{(n)}(x)$ at $x=\pm x_{\rm max}$, are fixed to $u^{(n)}(\pm x_{\rm max})=u_0^{(n)}(\pm x_{\rm max})$. See the appendix of ref.~\cite{Johnson:2020exp} for detailed tips on how to numerically solve the equations, and to solve the spectral problem for ${\cal H}[u]$ once $u(x)$ is found. Throughout, the choice $\hbar=1$ was made, although explorations were done with  $\hbar<1$, with qualitatively similar results.\label{foot}}

After finding an accurate enough numerical solution for $u(x)$, the spectrum of  the Schr\"odinger Hamiltonian $\mathcal{H}[u]$ of equation~(\ref{eq:68}) can be computed (again, using numerical techniques),  and ultimately the full spectral density $\rho(E)$ computed using (\ref{eq:69}).\footnote{The same numerical  techniques as the ones developed in refs.~\cite{Johnson:2019eik,Johnson:2020heh,Johnson:2020exp} were used, and so they will not be described here. See an appendix of ref.~\cite{Johnson:2020exp} for suggestions about solving the equations and extracting the spectrum of ${\cal H}$.} Picking the values $(\alpha_1,\alpha_2)=(1/4,1/3)$, the next step is to  analyse each sign of $\lambda$ separately.

\begin{figure}
\centering
\begin{subfigure}{0.47\textwidth}
\includegraphics[width=\textwidth]{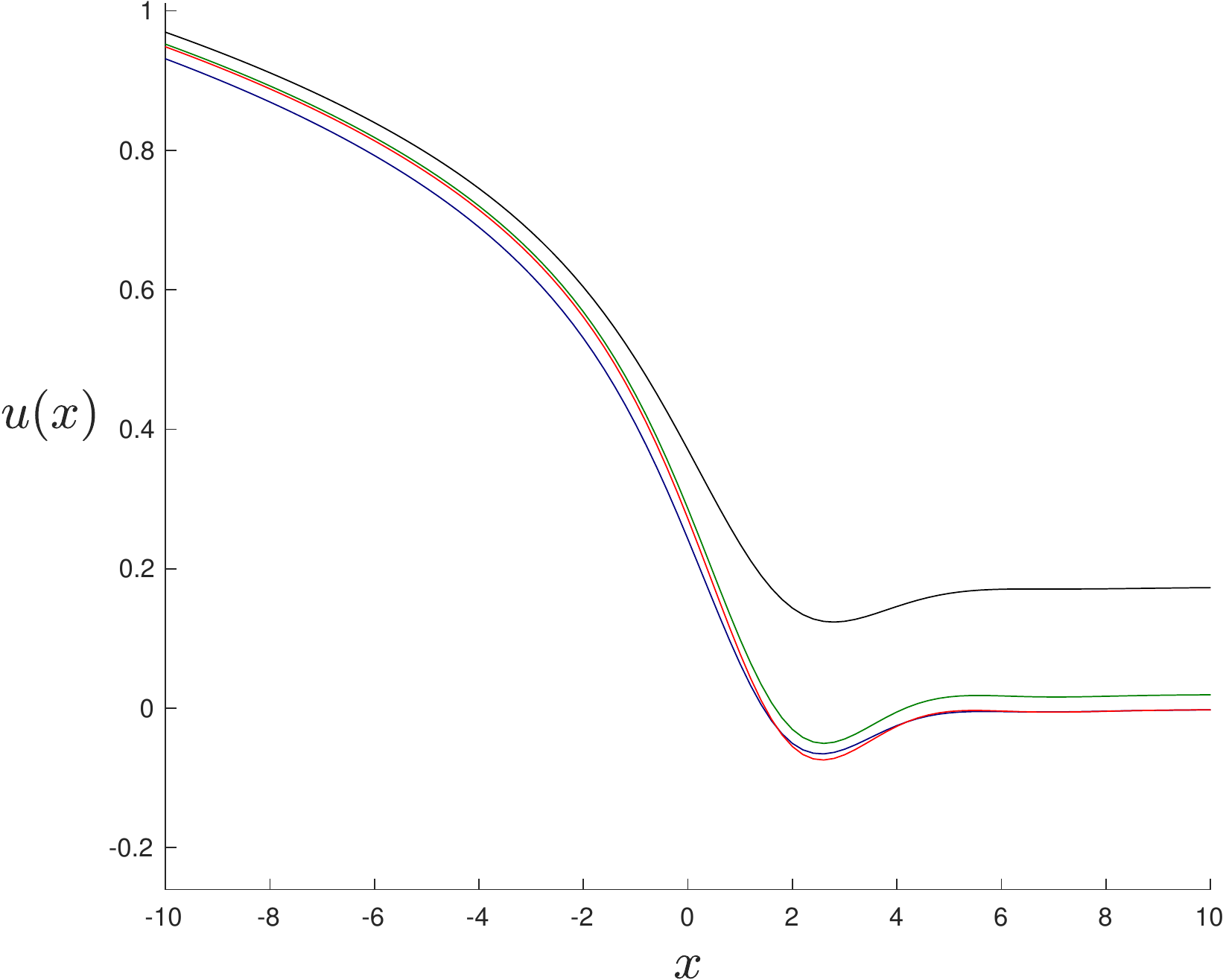}
\end{subfigure}
\begin{subfigure}{0.48\textwidth}
\includegraphics[width=\textwidth]{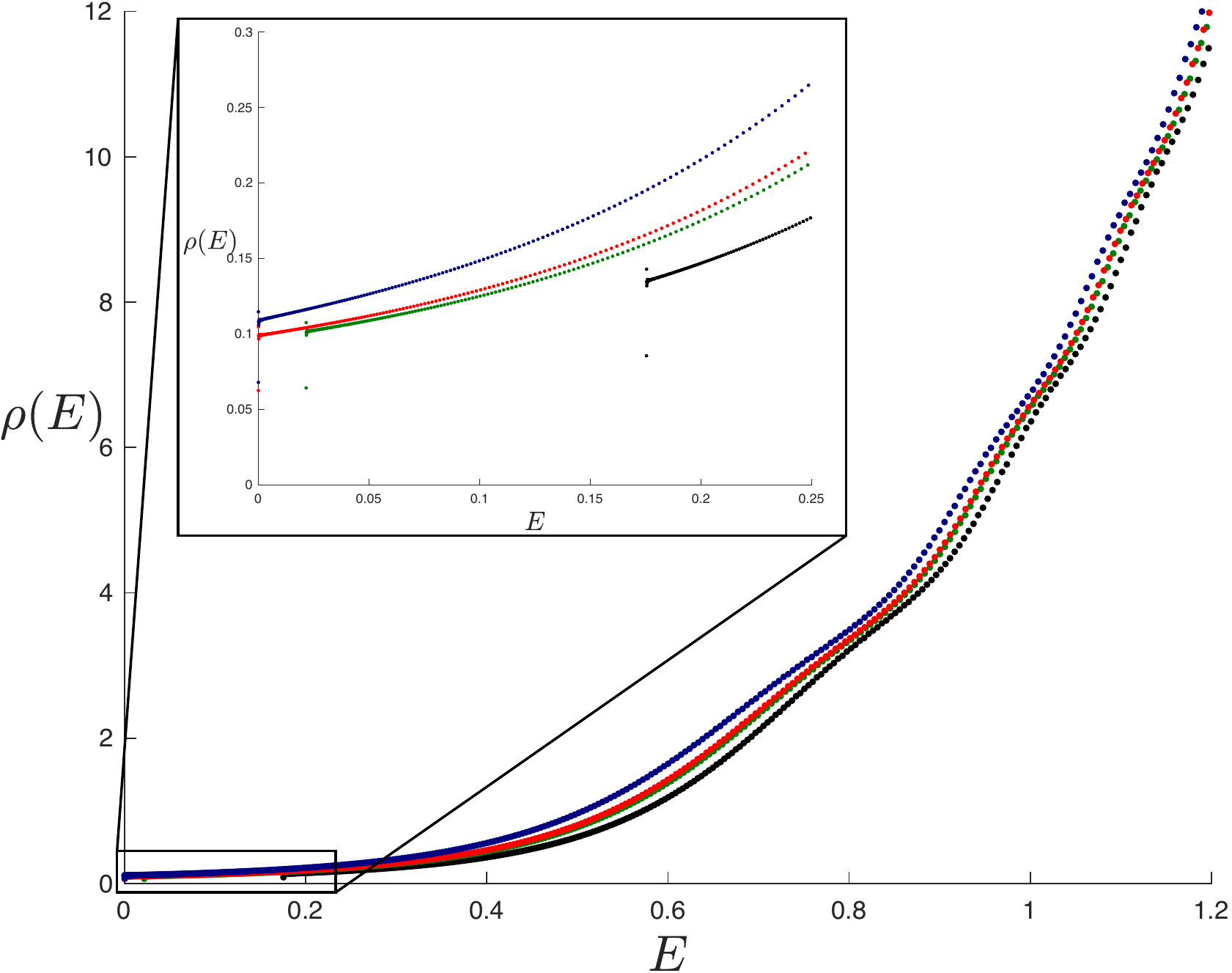}
\end{subfigure}
\caption{On the left are plots of the non-perturbative potential $u(x)$ in the $k_{\rm max}=7$ truncation for several $\lambda\ge 0$ and $(\alpha_1,\alpha_2)=(1/4,1/3)$. While the blue curve corresponds to pure JT gravity, red and green correspond to $\lambda=9\lambda_c/10$ and $\lambda= 11\lambda_c/10$ respectively, where $\lambda_c\simeq 1.042$ (\ref{eq:54}). The black curve is for $\lambda=2\lambda_c$. On the right (using the corresponding colours) are plotted the full spectral densities obtained from equation~(\ref{eq:69}) after computing the spectrum of $\mathcal{H}[u]$.}\label{fig:15}
\end{figure}

\paragraph{Positive $\lambda$:} The potential $u(x)$ was found, and the spectrum computed to enable the construction of $\rho(E)$ for positive $\lambda$ in the~${k_{\rm max}=7}$ truncation. 
In the left diagram of figure \ref{fig:15} is  plotted the  potential $u(x)$ for several values of $\lambda$, where the phase transition~(\ref{eq:transition_A}) in $E_0$ is manifest in the shift of the boundary condition  $u(+\infty)=E_0$. Comparing with the ordinary JT gravity solution with $\lambda=0$ (blue curve), it can be appreciated how the potential $u(x)$ is deformed. 

On the right diagram of figure~\ref{fig:15} is plotted the spectral density $\rho(E)$ computed from each of the potentials. For $\lambda=0$ it is known that the threshold density value~$\rho(0)$ is non-zero due to non-perturbative effects~\cite{Johnson:2019eik,Johnson:2020exp}. While for~$\lambda\in(0,\lambda_c)$ we also observe $\rho(0)$ non-zero (the slight numerical jitter near the threshold energy can be safely ignored), it takes a smaller value as compared to that of ordinary JT gravity (blue and red curves). As~$\lambda$ increases past the transition value $\lambda_c$, the threshold energy $E_0$ shifts to the  positive values obtained from equation~(\ref{eq:30}). The value of $\rho(E_0)$ is still non-zero and increasing with $\lambda$ (compare green and black curves).

The behavior of $\rho(E_0)$ is related to the value of the non-perturbative potential $u(x)$ at $x=0$. The spectral density $\rho(E)$ is computed from (\ref{eq:69}) by integrating over the (modulus-squared) tail of the $E=E_0$ wavefunction that penetrates into the~$x<0$ region. How much of that tail penetrates depends upon  the value of $u(x)-E_0$ at $x=0$. Here, note that $u(x)$ is  the {\it full} non-perturbative potential---its leading behavior $u_0(x)-E_0$ of course vanishes at $x=0$ by construction. Comparing the diagrams in figure \ref{fig:15} we observe that $\rho(E_0)$ increases whenever $u(x)-E_0$ decreases, since $|\psi(x,E_0)|^2$ (integrated to find $\rho(E_0)$ in (\ref{eq:69})) has a larger support in the region $x<0$.

\begin{figure}
\centering
\begin{subfigure}{0.47\textwidth}
\includegraphics[width=\textwidth]{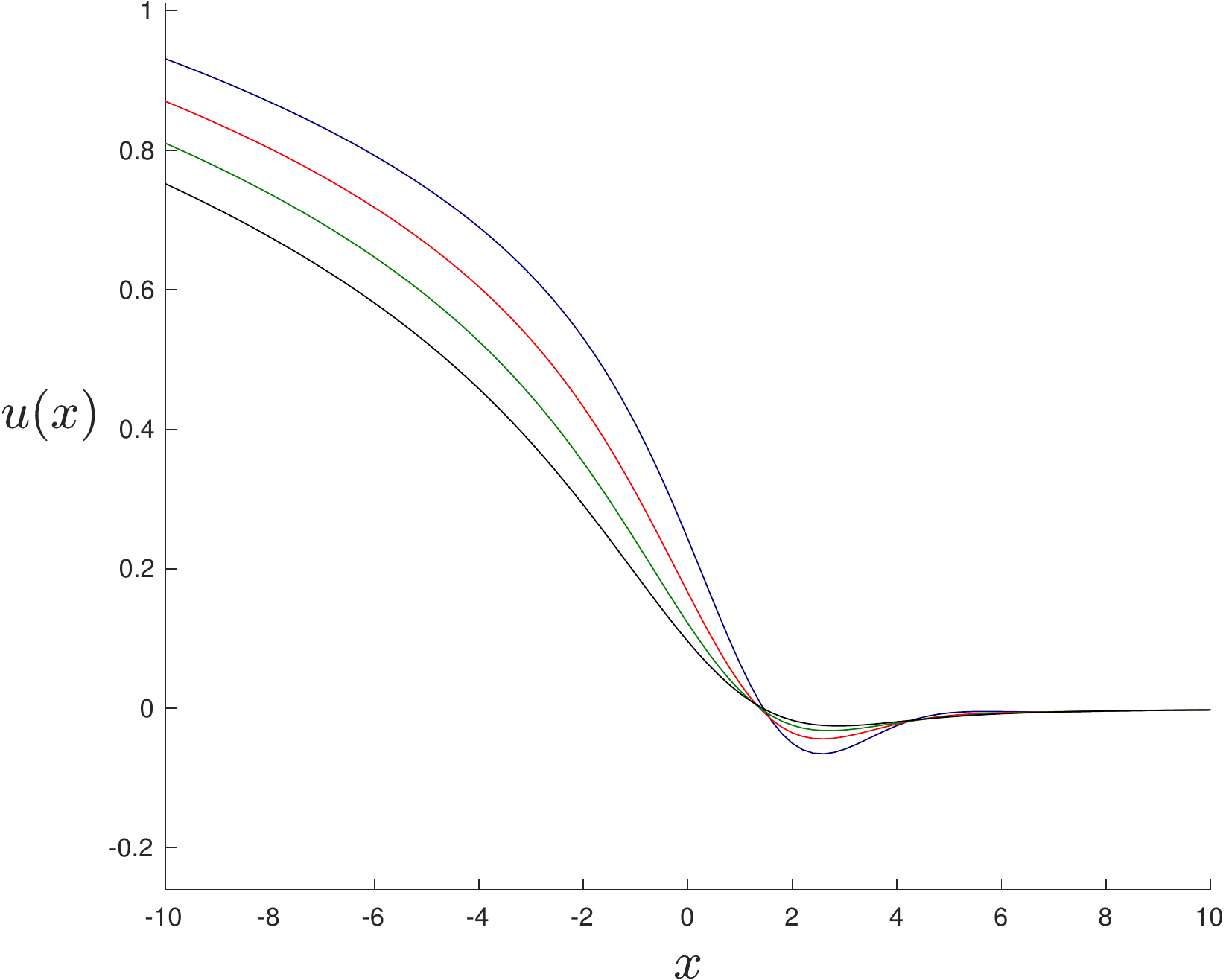}
\end{subfigure}
\begin{subfigure}{0.48\textwidth}
\includegraphics[width=\textwidth]{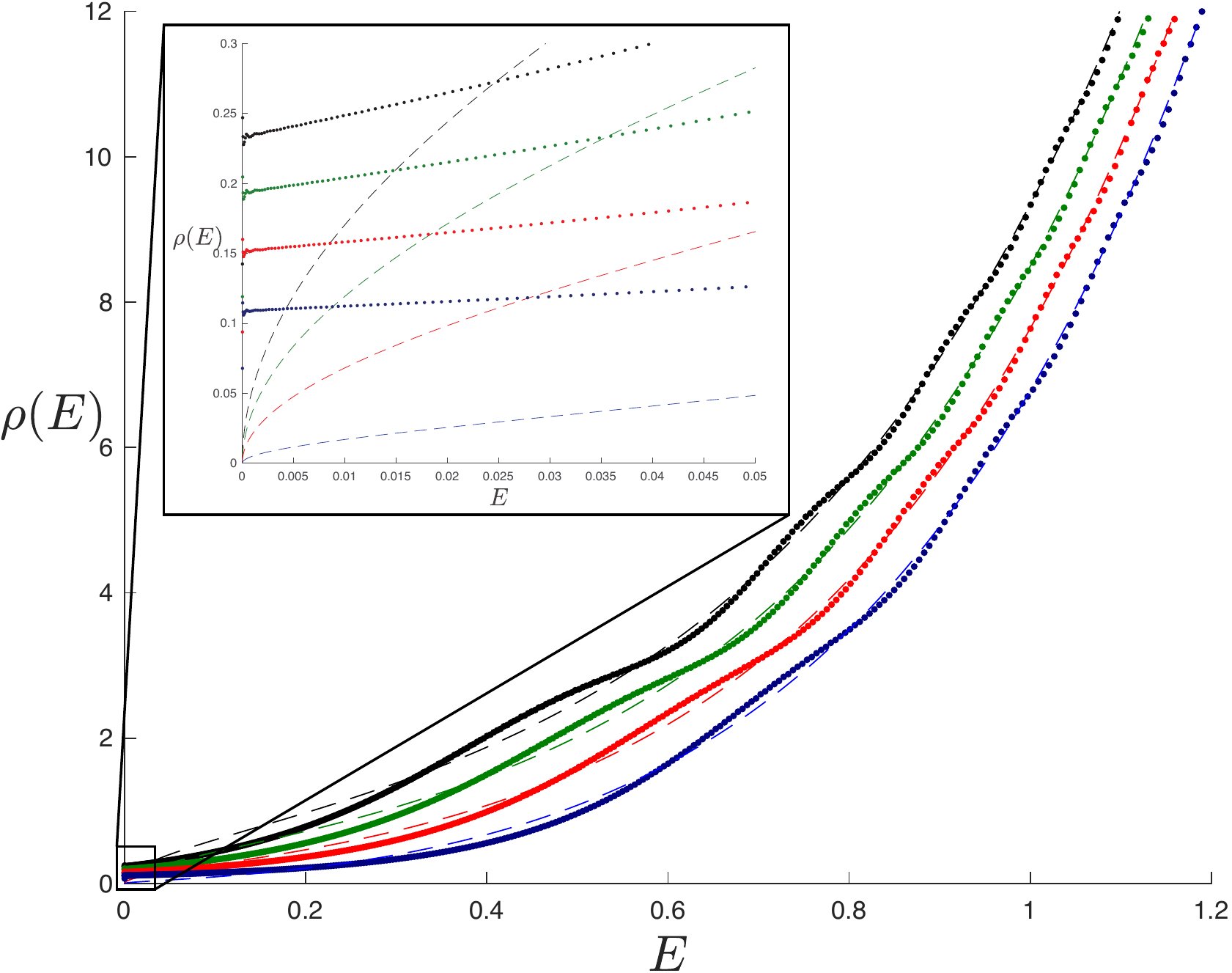}
\end{subfigure}
\caption{On the left are  plots of the non-perturbative potential $u(x)$ obtained from numerically solving~(\ref{eq:13}) for $(\alpha_1,\alpha_2)=(1/4,1/3)$ with $\lambda=0$ (blue), $\lambda=-10/3$ (red), $\lambda=-20/3$ (green) and $\lambda=-10$ (black). On the right are the associated spectral densities obtained from equation~(\ref{eq:69}) and plotted using the same colours, with the dashed lines corresponding to the disc spectral density $\rho_0(E)$ in (\ref{eq:41}).}\label{fig:14}
\end{figure}

\paragraph{Negative $\lambda$:} Let us now consider negative~$\lambda$, where there is no phase transition in the threshold energy~$E_0=0$. On the diagrams in figure~\ref{fig:14} we plot the non-perturbative numerical solutions to~$u(x)$ and~$\rho(E)$ for several values of~$\lambda$. The quantity $\rho(0)$, the spectral density at the threshold energy~$E_0=0$,  increases for increasingly larger negative~$\lambda$. From the discussion immediately above, this is in agreement with the value of the potential $u(x)$ at the origin~$x=0$.

Overall, we conclude that the deformation of JT gravity according to (\ref{eq:70}) is both perturbatively and non-perturbatively well defined (using the non-perturbative scheme adopted in this paper). For $\lambda$ positive the phase structure obtained from the semi-classical analysis of the two-dimensional black holes solutions is in qualitative agreement with the full partition function as captured by the matrix model. For~$\lambda$ negative (we explored out to $\lambda=-100$ to ensure that the patterns of figure~\ref{fig:14} persist) this does not appear to be the case, as the energy gap~$E_{\rm gap}$ observed in the semi-classical approximation is invisible in the matrix model description, even after accounting for non-perturbative effects.\footnote{The first version of this manuscript reported effects that were interpreted as the appearance of a non-perturbative~$E_{\rm gap}$. They were due to a subtle error in the  code used to numerically solve the full string equation. This error has been corrected.}

\section{Deformed JT Gravity: Model B}
\label{sec:4}

This section will repeat the previous the analysis of Euclidean partition function but for a different deformation of JT gravity, given by:
\begin{equation}\label{eq:63}
U(\phi)=2\lambda e^{-2\pi(1-\alpha)\phi}\ ,
\end{equation}
where now $U(0)\neq 0$. We show that for certain values of $\lambda$ this deformation is both perturbatively and non-perturbatively inconsistent, using our definition in section~\ref{sec:non-perturbative}.

\subsection{Semi-Classical Approximation}
\label{sub:4.1}

As done in section~\ref{sub:3.1}, the semi-classical approximation can be used to compute the Euclidean partition function~(\ref{eq:37}) of the theory with potential $W(\phi)=2\phi+U(\phi)$, but now with $U(\phi)$ given in equation~(\ref{eq:63}). The black hole solutions are written in equation~(\ref{eq:62}) and depend on the single parameter $\phi_h$. The conditions derived on $W(\phi)$ have the same interpretation as before, and the thermodynamic quantities are computed from equation~(\ref{eq:53}).\footnote{A semi-classical analysis of this theory was recently presented in appendix D of ref.~\cite{Maxfield:2020ale}. There is a slight difference in the proportionally factor in $U(\phi)$ that does not modify the qualitative behavior.}

\begin{figure}
\centering
\includegraphics[scale=0.34]{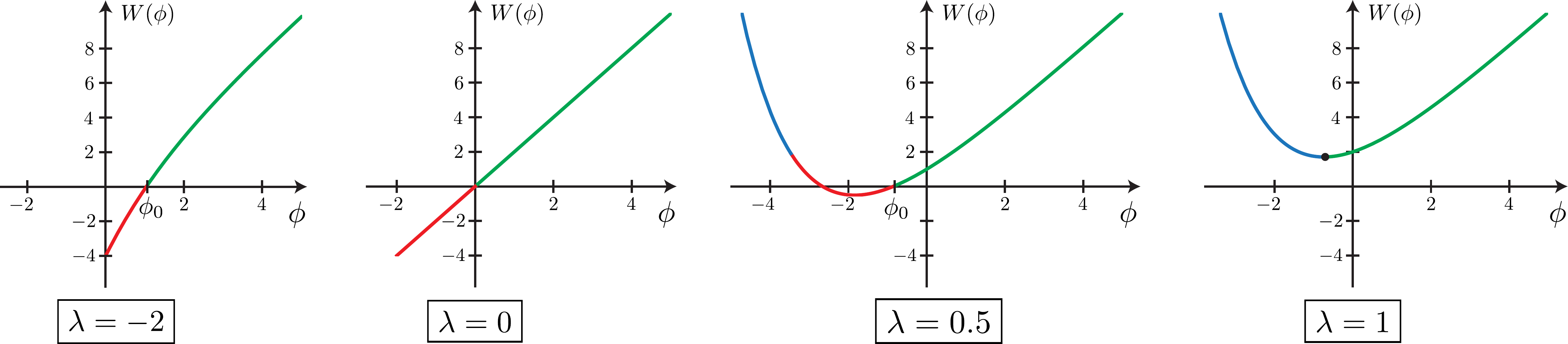}
\caption{Deformed JT gravity dilaton potential $W(\phi)=2\phi+U(\phi)$ in (\ref{eq:63}) for $\alpha=0.9\sim 1$ and several values of $\lambda$. Black hole solutions exist for $\phi=\phi_h$ indicated in green (stable) and blue (unstable). For $\lambda>\lambda_c$ in (\ref{eq:74}) there is a transition to a regime in which there are no zero temperature black hole solutions.}\label{fig:11}
\end{figure}

In figure \ref{fig:11} the potential is plotted for  $\alpha=0.9\sim 1$ and several values of $\lambda$ in order to  illustrate its key qualitative features. As before, segments of the curve in red, blue and green indicate whether the black hole solution with $\phi_h=\phi$ is non-existent, unstable or stable. For $\lambda$ negative there is a single branch of stable black hole solutions, with the zero temperature solution determined by~${\phi_h=\phi_0(\lambda)>0}$. The value of $\phi_0(\lambda)$ changes smoothly with $\lambda$ and there are no finite temperature phase transitions. For positive $\lambda$ there are two separate branches with stable and unstable black holes. At $\lambda=\lambda_c>0$ there is a  transition to a phase for which there is no zero temperature black hole solution. The critical value associated to this transition can be easily computed and is given by
\begin{equation}\label{eq:74}
W(\phi_0^{(c)})\big|_{\lambda=\lambda_c}=
W'(\phi_0^{(c)})\big|_{\lambda=\lambda_c}=0\ ,
\qquad \qquad
\lambda_c=\frac{1}{2\pi e(1-\alpha)}\ ,
\qquad
\phi_0^{(c)}=
\frac{-1}{2\pi(1-\alpha)}
\ ,
\end{equation} where $e$ is Euler's number.
Note that this is a different kind of transition than the one analysed around equation~(\ref{eq:51}). While in that case the transition was between two different zero temperature regimes, in this case there is a transition between a system with a $T=0$ regime and one without.

\subsection{Perturbative Genus Expansion}
\label{sub:4.2}

\begin{figure}
\centering
\includegraphics[scale=0.30]{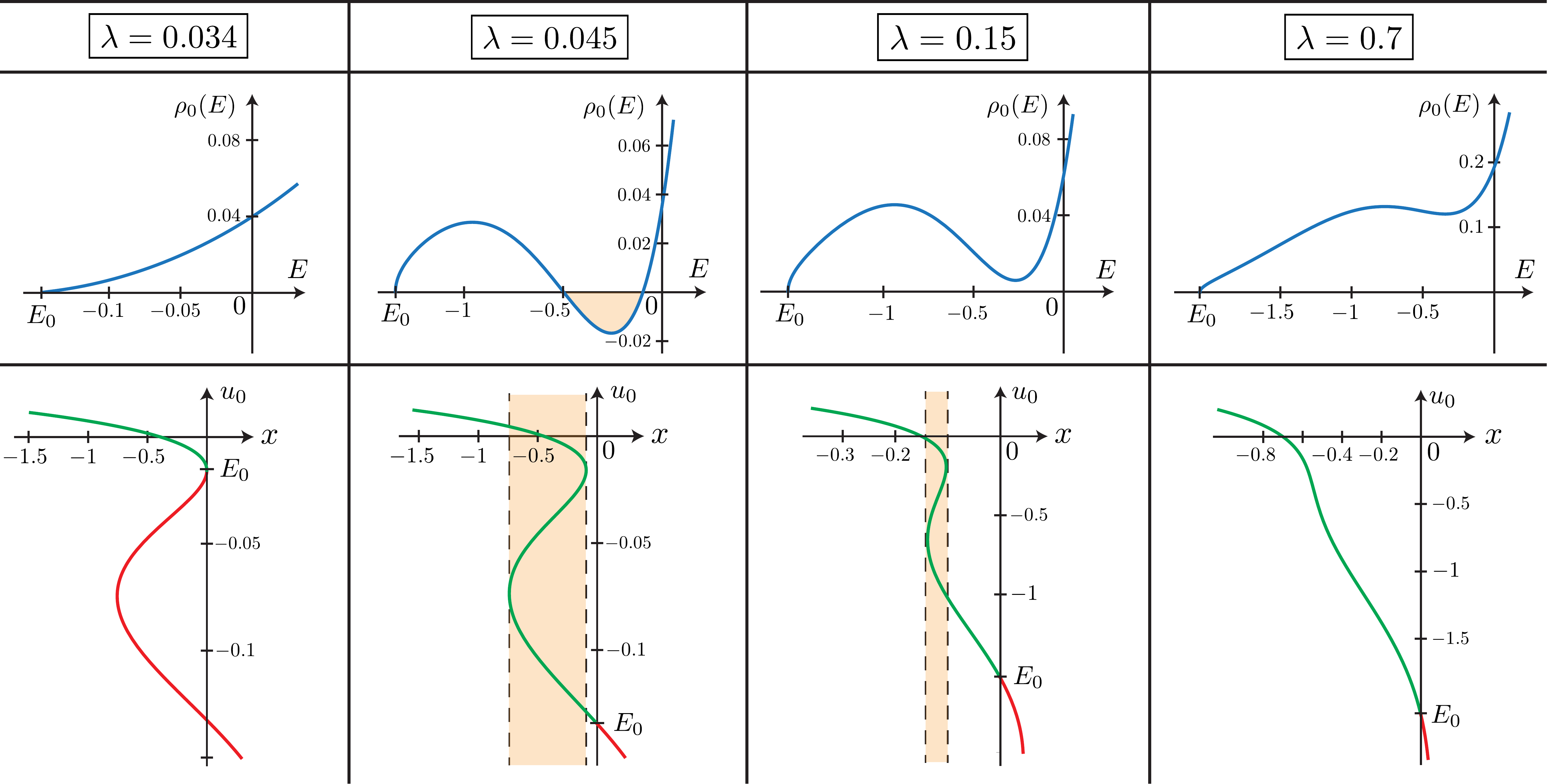}
\caption{The first row corresponds to the the $\rho_0(E)$ numerically computed from (\ref{eq:45}) for $\alpha=1/4$ and several values of $\lambda$. For $\lambda\in(0.035,0.122)$ the spectral density is non-positive. In the second row we plot the curve $x=x(u_0)$ in the $(x,u_0)$ plane obtained from (\ref{eq:55}) with $\alpha=1/4$ and the same values of $\lambda$. The function $u_0(x)$ is multi-valued in the shaded region in $x<0$ for a (different) finite window of~$\lambda$, approximately given by $\lambda\in(0.035,0.334)$.}\label{fig:16}
\end{figure}

Consider now the matrix model computation of the partition function at leading (disc) order, as in equation~(\ref{eq:73}). The spectral density computed in refs.~\cite{Maxfield:2020ale,Witten:2020wvy} is obtained from the integral~(\ref{eq:45}), where $E_0$ is obtained from $\mathcal{R}_0[E_0,0]=0$ with
\begin{equation}\label{eq:55}
\mathcal{R}_0[u_0,x]=
\frac{\sqrt{u_0}}{2\pi}
I_1(2\pi\sqrt{u_0})+
\lambda\, I_0(2\pi\alpha\sqrt{u_0} )+x=0\ .
\end{equation}
The top row of figure \ref{fig:16} displays the  spectral density $\rho_0(E)$ (extracted numerically from~(\ref{eq:45}) with~(\ref{eq:55})) for $\alpha=1/4$ and several values of $\lambda$. Strikingly, $\rho_0(E)$ develops negative regions for a finite window of $\lambda$ approximately given by $\lambda\in(0.035,0.122)$. As discussed before, this clearly signals a problem, and again its origins lie in the behavior of $u_0(x)$ as a solution to equation~(\ref{eq:55}). 

The place to look at is the implicit definition of $u_0(x)$ from the string equation~(\ref{eq:55}). In the second row of figure~\ref{fig:16} is plotted the curve $x=x(u_0)$ for  the same values of $\alpha$ and $\lambda$ as before. From the first two diagrams starting from the left, we note that after $\lambda\simeq 0.034$, the threshold energy $E_0$ jumps discontinuously as a function of~$\lambda$. This zeroth order phase transition generates a multi-valued potential $u_0(x)$ in the crucial region $x<0$ as can be seen at {\it e.g.,}  $\lambda=0.045$. This ultimately results (for the reasons discussed after equation~(\ref{eq:25})) in a negative spectral density. Amusingly, while a phase transition in $E_0$ solved the issues of $\rho_0(E)$  in the previous section, in this case it is the origin of the problem.

Since the potential $u_0(x)$ is multivalued in the region $x<0$, the issue is more severe than the one discussed in the previous section. This is because leading order observables in the matrix model are computed from the behavior of $u_0(x)$ in $x<0$. Recall for instance, the disc partition function~$Z_0(\beta)$, given in terms of~$u_0(x)$ in equation~(\ref{eq:56}).
That expression makes no sense if the function $u_0(x)$ is multivalued in the region $x<0$, meaning the~${\hbar\rightarrow 0}$ limit that defines~$Z_0(\beta)$ does not exist. Since any single trace observable (like the spectral density) can be computed from~$Z_0(\beta)$, it means none of these observables have a well defined leading genus behavior. This results in the breakdown of the~$\hbar$ perturbative expansion of the matrix model.\footnote{Moreover, from the point of view of the associated quantum mechanical problem,  having a multi-valued potential for $u_0(x)$, in the classical limit or beyond, is highly problematic.} This type of failure is not extremely unusual, for instance observables in a multi-cut Hermitian matrix model do not have the ordinary power series expansion in $1/N$ \cite{Bonnet:2000dz}.

It must be emphasized that the multi-valuedness of  $u_0(x)$ for $x<0$ is the {\it root cause} of the physical problems here. Negativity of $\rho_0(E)$ is only a symptom. A more careful comparison of the diagrams in figure \ref{fig:16} illustrates this. Since for~${\lambda=0.15}$ the leading genus spectral density is positive, one might think the matrix model perturbative expansion in this case is sensible. However, a closer inspection shows that $u_0(x)$ for $\lambda=0.15$ is still multi-valued in the region $x<0$, meaning the leading genus spectral density obtained from equation~(\ref{eq:56}) is not defined at all. This shows the negativity of the spectral density is not a robust indicator when assessing the health of the matrix model.

\subsection{Non-Perturbative Effects}
\label{sub:4.3}

The breakdown of the perturbative expansion calls for an appeal to non-perturbative physics. Here again, the definition of section~\ref{sec:non-perturbative} is employed. To do so, the coefficients $t_k$ appearing in~$\mathcal{R}[u,x]$ in equation~(\ref{eq:77}) must be obtained. These can be identified by expanding the disc level equation~(\ref{eq:55}), giving:
\begin{equation}\label{eq:58}
\mathcal{R}_0[u_0,x]=
\sum_{k=1}^{\infty}t_k(\lambda)u_0^k+\lambda+x\ ,
\qquad {\rm where} \qquad
t_k(\lambda)=\frac{\pi^{2(k-1)}}{2k!^2}
\big(
k+2\pi^2\lambda \alpha^{2k}
\big)\ .
\end{equation}
Using these coefficients we can write $\mathcal{R}[u,x]$ and the non-perturbative differential equation (\ref{eq:13}) for~$u(x)$. As discussed in the case of section~\ref{sub:3.3}, explicit solution can be obtained {\it via} a truncation  to some highest $k=k_{\rm max}$ as in equation~(\ref{eq:21}). It is important to pick a truncation that captures the phase transition of $E_0$ shown in figure \ref{fig:16}. Only odd truncations $k_{\rm max}$ are useful in this regard, with~$k_{\rm max}=5$ the minimum truncation with the necessary structure, given the properties of the~$t_k(\lambda)$. To obtain a more accurate spectral density for a larger range of energies, the truncation~$k_{\rm max}=7$ is used. The curve~$x=x(u_0)$ obtained from~$\mathcal{R}^{(7)}_0[u_0,x]=0$ is analogous to the ones shown in figure~\ref{fig:16}, with the transition occurring at $\lambda_c\simeq 0.03480$. For $k_{\rm max}=7$, the differential equation is $14$th order  but as before (see footnote~\ref{foot}) some persistence can yield highly accurate numerical solutions for~$u(x)$ (errors smaller than~${10^{-4}}$).

\begin{figure}
\centering
\begin{subfigure}{0.48\textwidth}
\includegraphics[width=\textwidth]{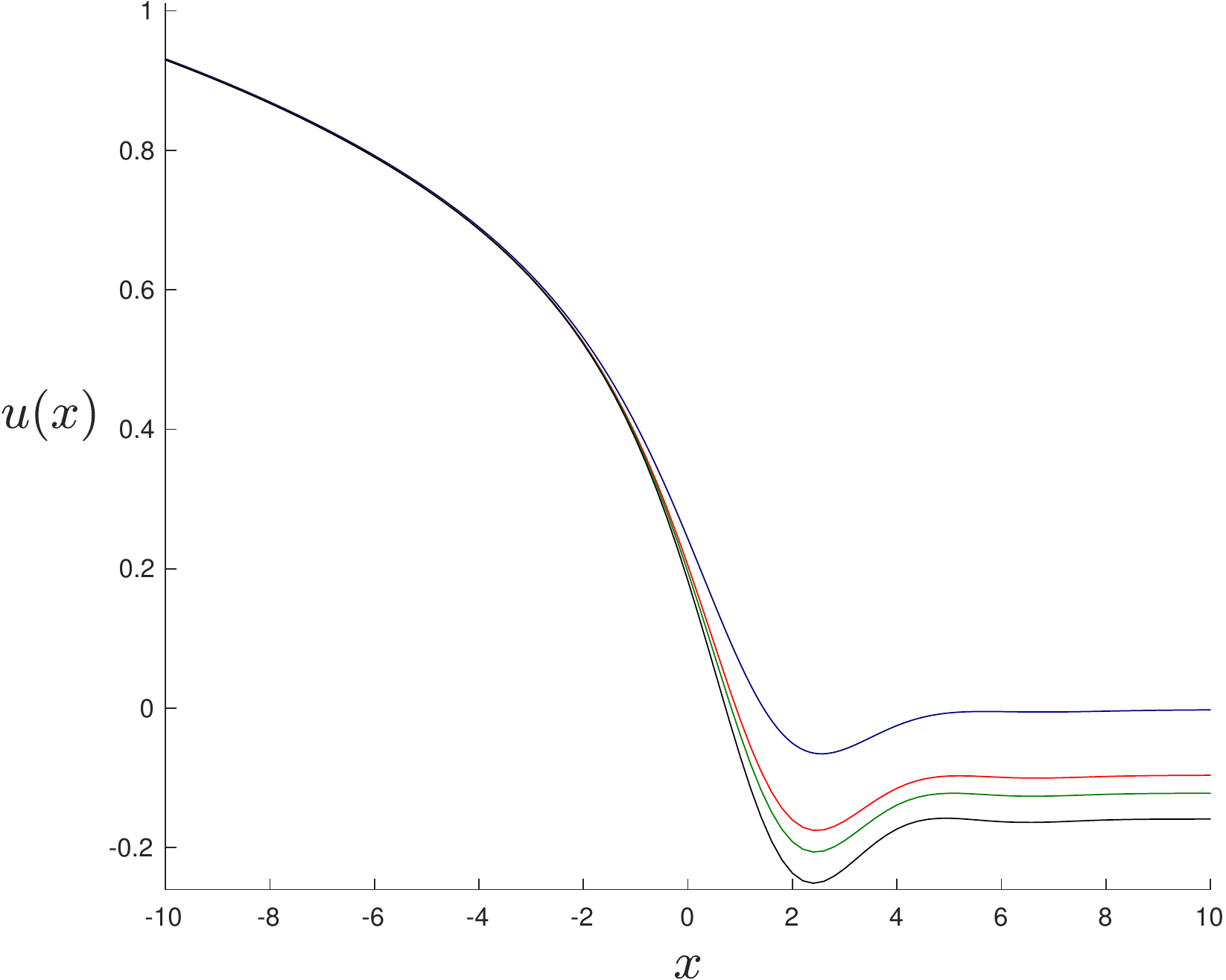}
\end{subfigure}
\begin{subfigure}{0.50\textwidth}
\includegraphics[width=\textwidth]{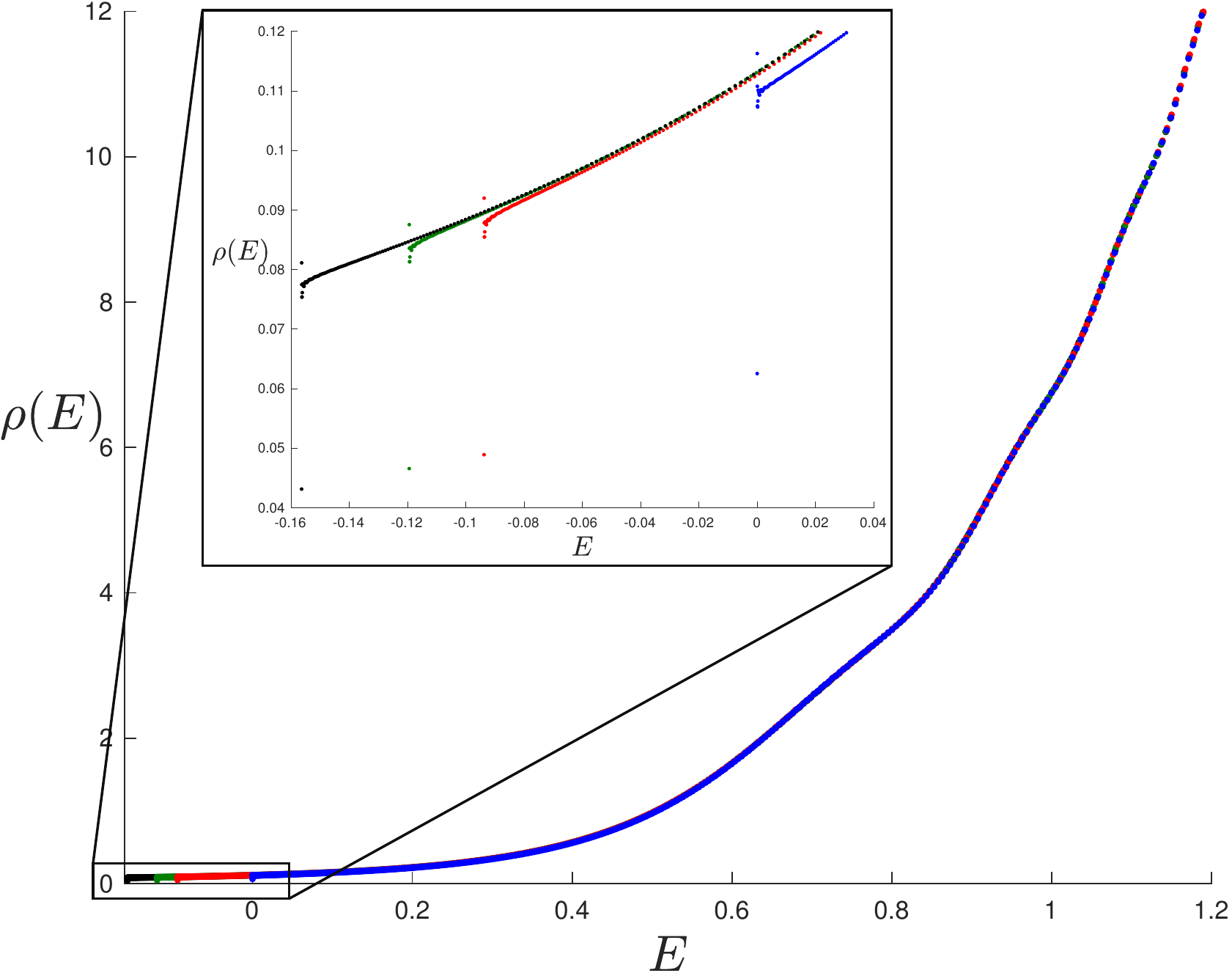}
\end{subfigure}
\caption{On the left are numerical solutions for the potential $u(x)$ obtained by solving  the  string equation~(\ref{eq:13}) for $\alpha=1/4$ in the $k_{\rm max}=~7$ truncation and $\lambda=0$ (blue), $\lambda= 0.03$ (red), $\lambda= 0.033199$ (green) and $\lambda=0.034798$ (black). All of these values are before the zeroth order transition of $E_0$ that occurs at $\lambda_c\simeq 0.03480$.  On the right are the associated spectral densities obtained from (\ref{eq:69}) and plotted using the same colors.}\label{fig:8}
\end{figure}

\paragraph{Before the transition:} First is the case of $\lambda<\lambda_c$, where~$\lambda_c$ is the critical value triggering the transition in $E_0$. Similarly to the previous section solutions can be found for the non-perturbative potential $u(x)$ and its associated spectral density $\rho(E)$ computed. The results are shown in figure \ref{fig:8} for several values of $\lambda<\lambda_c$. There is no phase transition for $E_0$, and its value changes continuously with $\lambda$, as can be seen from the behavior at $u(+\infty)=E_0$. The spectral density $\rho(E)$ does not show any particularly novel behavior for $\lambda<\lambda_c$ (in contrast to some of the phenomena seen in the previous section).

\paragraph{Across and beyond the transition:} Next is a study of the non-perturbative behavior of the model as $\lambda$ crosses the transition value $\lambda_c$ where $E_0$ has a finite jump. In the $k_{\rm max}=7$ truncation the range of $\lambda$ for which the perturbative expansion breaks down is given by $\lambda\in(0.03480,0.36733)$, with the left edge corresponding to the zeroth order phase transition in $E_0$. 

As soon as $\lambda$ crosses the transition the string equation (\ref{eq:13}) for $u(x)$ becomes extremely unstable. While for $\lambda$ arbitrarily close to $\lambda_c$ before the transition, well behaved solutions can be constructed, the numerics becomes unpredictable as soon as $\lambda$ goes beyond $\lambda_c$.\footnote{Numerical instability can manifest itself in a number of ways. The clearest and most reliable sign of instability, seen here, is when inaccurate solutions are found (sometimes with sudden increase in run time for convergence to occur), and for which there is strong dependence in their shape to small changes in the size of the grid, and its granularity.} Before jumping to the conclusion that the deformed JT gravity theory is non-perturbatively unstable beyond the transition, it is prudent to carefully consider other alternatives that might explain the issue. The first obvious possibility is a lack of numerical precision: the discontinuous jump in $E_0$ might be making the differential equation harder to solve, even though a well behaved solution might in principle still exist. Deforming away from a neighboring solution already found (a common trick for numerically exploring difficult equations) does not work here since $E_0$ jumps.  To explore this, a toy model  was constructed that has the same features as the deformed JT gravity theory with the difference that   the jump in $E_0$ can be controlled and made small. The model  has the following leading genus string equation:
\begin{equation}\label{eq:61}
\mathcal{R}_{0}^{\rm toy}[u_0,x]=
u_0^2(u_0+a)(u_0^2+u_0+1)+b+x=0\ ,
\end{equation}
where $b\in \mathbb{R}$ and $a>0$. While for $b=0$ this model has $u_0(0)=E_0=0$, a zeroth order phase transition is triggered when $b>0$, where the threshold energy jumps to $E_0\simeq a$ and~$u_0(x)$ becomes multi-valued in the region $x<0$. The plot of~$u_0(x)$ for this toy model is analogous to the first two plots in the second row of figure~\ref{fig:16}. By tuning the parameter $a$, the discontinuity in $E_0$ across the phase transition can be  made as small as needed. The full  non-perturbative string equation~(\ref{eq:13}) associated to this toy model is written after identifying the coefficients $t_k(a,b)$ from (\ref{eq:61}). Attempting to numerically solve the resulting equation for $b>0$, we find the instabilities are still present, no matter how small the parameter $a$ is taken to be. This shows the instability issue is not addressed by reducing the magnitude of the jump in $E_0$.

\begin{figure}
\centering
\includegraphics[scale=0.40]{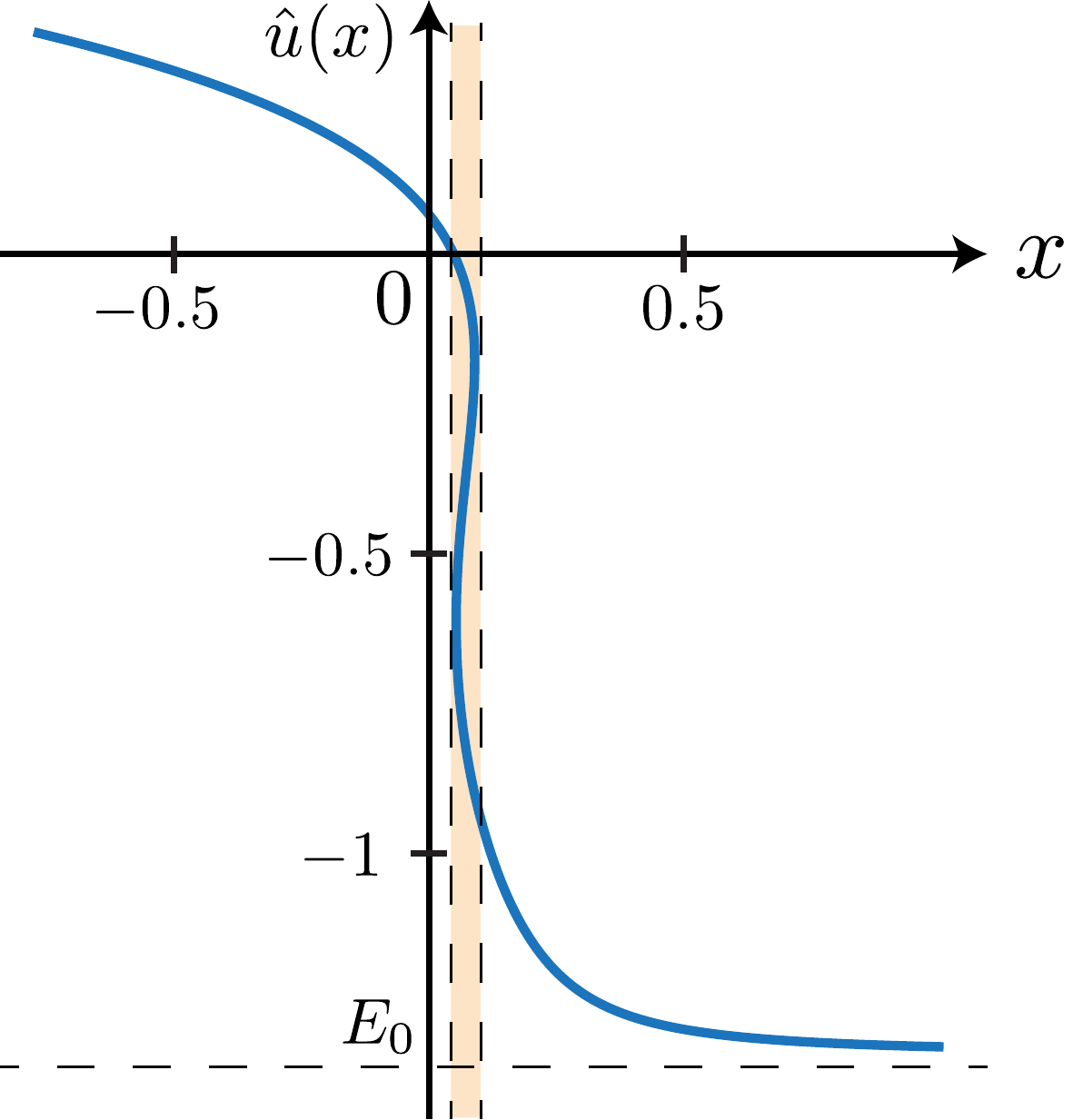}
\caption{Numerical plots obtained by solving for the potential $\hat{u}(x)$ in (\ref{eq:22}) corresponding to the t'Hooft limit with $q=0.1$ and $\lambda=0.038>\lambda_c=0.035$. The asymptotic behavior of the solutions shows the distinct behavior at $\hat{u}(+\infty)=E_0$. There is a shaded region for which $\hat{u}(x)$ is multi-valued.}\label{fig:9}
\end{figure}  

There is an analytical argument that has been previously used in refs.~\cite{Klebanov:2003wg,Iyer:2010ex} to argue in favor of the existence of solutions to the differential equation (\ref{eq:13}) (and variants of it), when the right hand side has the  constant $\hbar^2\Gamma^2$ instead of 0. It involves taking a t'Hooft limit as $\hbar\to0$  such that~$\Gamma$ becomes large while the combination $q=\hbar\Gamma$ remains finite. In this limit, the string equation  simplifies to the following algebraic constraint:
\begin{equation}\label{eq:22}
(\hat{u}-E_0)\left(
\sum_{k=1}^{\infty}t_k(\lambda)\hat{u}^k+\lambda+x
\right)^2=q^2\ ,
\end{equation}
where $\hat{u}(x)\equiv \lim_{(\hbar,1/\Gamma)\rightarrow 0}u(x)$. Smooth algebraic solutions of this equation for $\hat{u}(x)$ suggest the existence of solutions $u(x)$ for other $\hbar$ and $\Gamma$. The algebraic constraint for $\hat{u}(x)$ in (\ref{eq:22}) can be easily solved numerically and visualized, see figure \ref{fig:9} for a plot for a fixed $q$ and $\lambda>\lambda_c$. The solution $\hat{u}(x)$ automatically satisfies the appropriate boundary conditions, and for large $q$ a smooth single-valued solution exists. However for small enough $q$ the solution $\hat{u}(x)$ is multi-valued within a certain range in $x$ (which interestingly occurs for $x>0$).  

The emergence of the multi-valuedness (for small enough $q$) suggests that there is no path connecting possible smooth stable solutions of the full differential equation that may exist at finite~$\Gamma$ to the ones needed at $\Gamma=0$.\footnote{Such a path can be found when there is no multi-valuedness, by for example using a solution-generating transformation~\cite{Dalley:1992br,Carlisle:2005mk} that acts on solutions of the string equation to generate new functions that are solutions of the string equation with  $\Gamma$ changed by an integer. It is in fact a kind of ``B\"acklund'' transformation, known in the integrable systems literature for changing  the soliton number of certain kinds  of solutions to integrable systems, here the relevant integrable system is KdV. The transformation can be made explicit, and an extension of it for cases with non-zero $E_0$ described and derived in Appendix~\ref{sec:Backlund}.} More generally, it is difficult to see how a multivalued solution of an algebraic equation, our  $u_0(x)$, (or even one smoothed out by the $q$-deformation above) can be completed into a solution of a boundary value problem for an ordinary differential equation once~$\hbar$ is turned on. The multi-valuedness overconstrains the information supplied by the boundary conditions. As a last remark, it is with noting that a multivalued classical potential $u_0(x)$ also makes no sense from the point of view of the quantum mechanical problem on the real line $x\in\mathbb{R}$. At a given point $x$ there would be an ambiguity as to the value of the force $F_x=-\partial_x u_0(x)$.

%
%

\begin{figure}
\centering
\begin{subfigure}{0.47\textwidth}
\includegraphics[width=\textwidth]{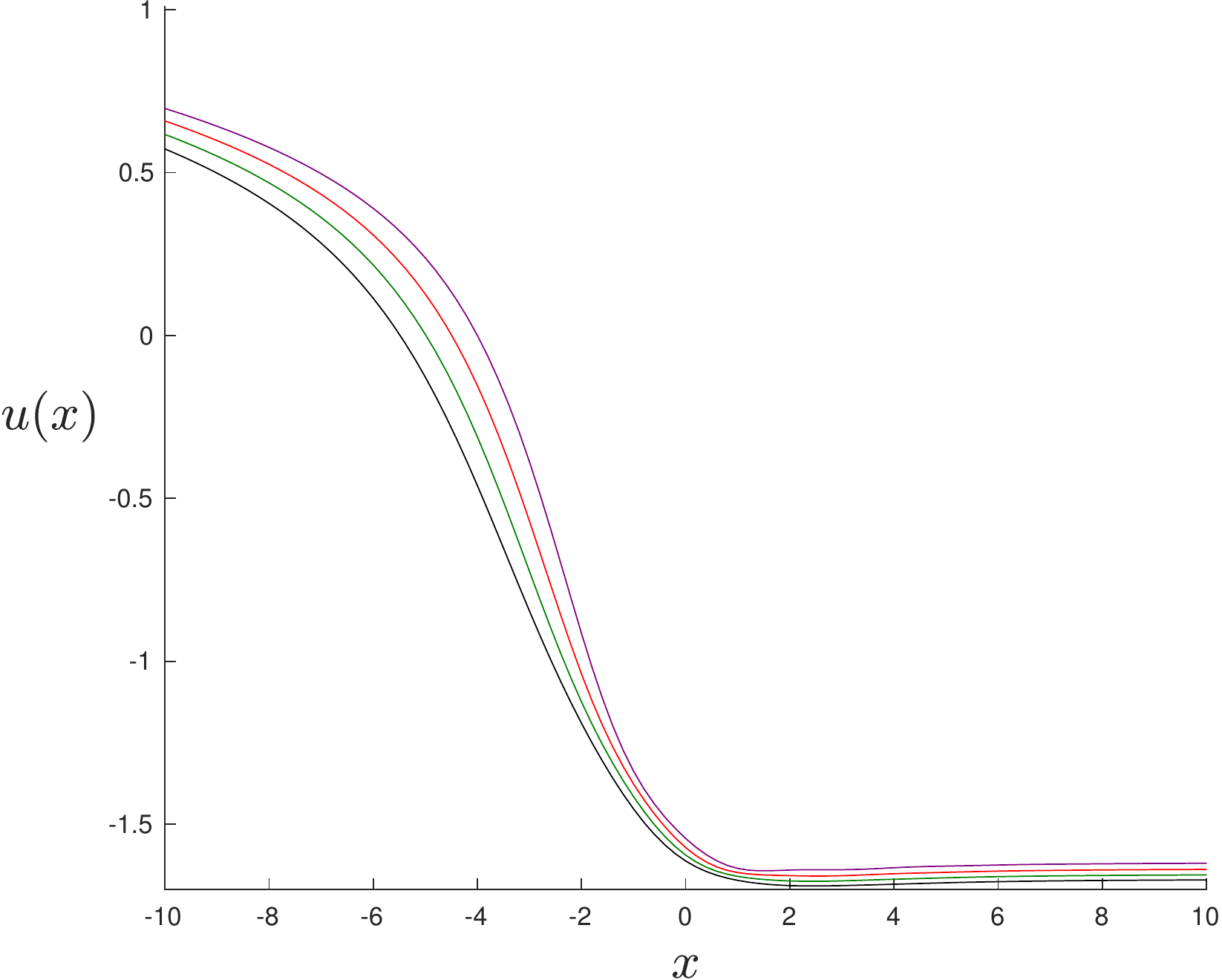}
\end{subfigure}
\begin{subfigure}{0.52\textwidth}
\includegraphics[width=\textwidth]{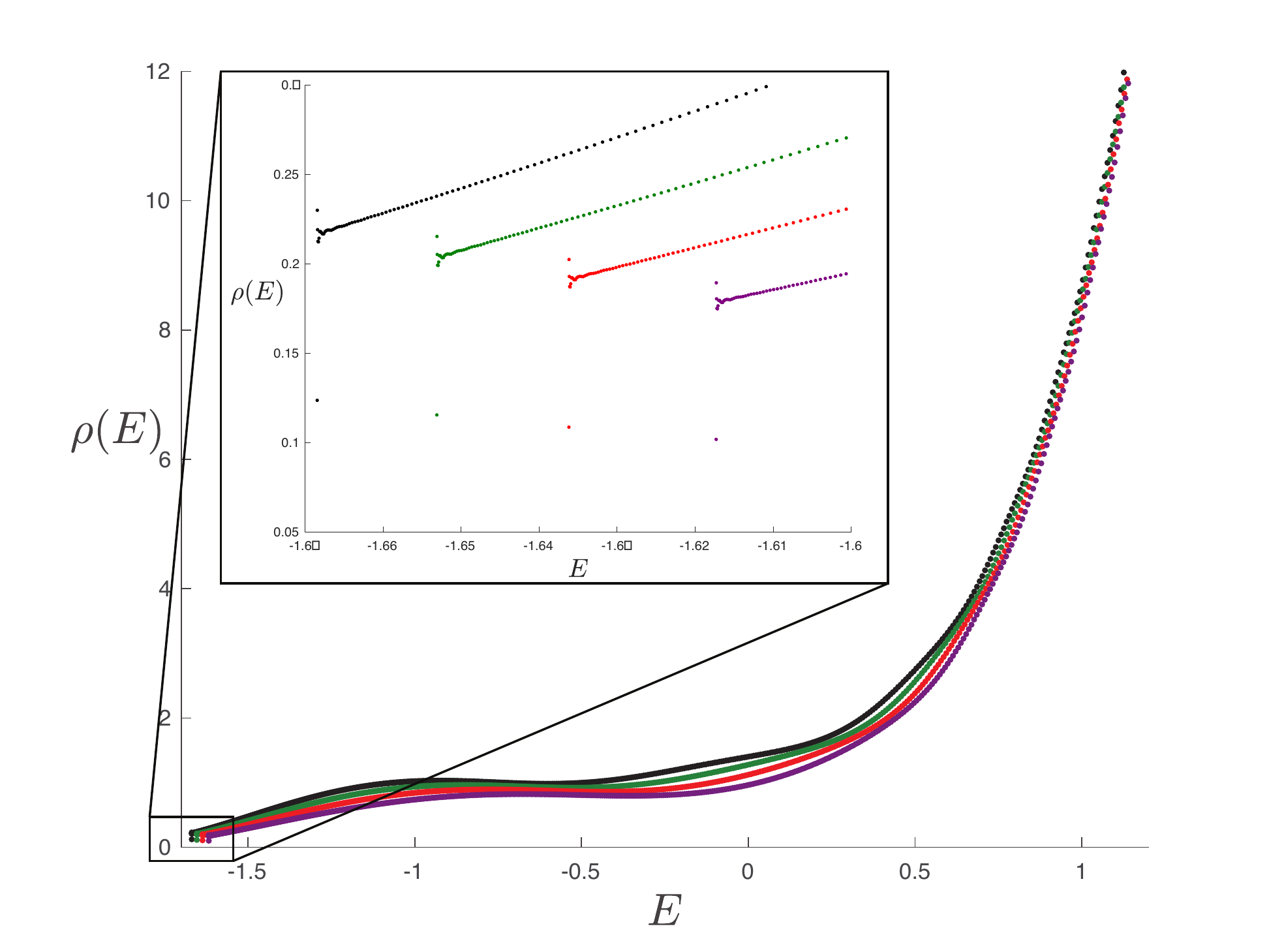}
\end{subfigure}
\caption{On the left are numerical solutions for the potential $u(x)$ obtained by solving  the  string equation~(\ref{eq:13}) for $\alpha=1/4$ in the $k_{\rm max}=~7$ truncation and $\lambda$ beyond the window $\lambda\in(0.03480,0.36733)$ in which $u_0(x)$ is multi-valued (see figure \ref{fig:16}). Here $\lambda=4$ (purple), $\lambda=4.5$ (red), $\lambda=5$ (green) and $\lambda=5.5$ (black). On the right  the associated spectral densities obtained from (\ref{eq:69}) and plotted using the same colors.
}\label{fig:12}
\end{figure}

All the evidence therefore points towards a non-perturbative instability or non-existence of the deformed JT gravity theory for~$\lambda$ across the transition, when~$u_0(x)$ becomes multi-valued in the~$x<0$ regime. Since~$u_0(x)$ is multi-valued in the finite window~$\lambda\in(0.03480,0.36733)$, this means we should be able to construct numerical solutions for $\lambda$ on the other side of this window, \textit{i.e.}~${\lambda>0.36733}$. Figure \ref{fig:12}  shows that this is indeed the case, displaying regular solutions to~$u(x)$ for~$\lambda$ beyond the multi-valued window of~$u_0(x)$. The associated spectral densities are also well-behaved and shown in that figure. Overall, there is enough evidence to reach the following conclusion about the leading~$u_0(x)$ (defined, possibly piecewise, as a continuous function over the whole domain~${-\infty\leq x \leq+\infty}$):
$$
\boxed{
{\rm For\,\,\lambda\,\,with\,\,}u_0(x)\,\,{\rm multivalued,\,\,
the\,\,system\,\,is\,\,non\mbox{-}perturbatively\,\,unstable}
}
$$
This is the central result of this section. 

It is worth remarking that  the above statement/result has a more general validity. In the case of $\lambda=0$, ordinary JT gravity, the non-perturbative string equation from a Hermitian matrix model definition is  just ${\cal R}=0$. There, the full $u_0(x)$ is multivalued in the $x>0$ regime, and the statement about multi-valuedness above is consistent with the fact that the string equation does not seem to have a stable solution. The non-perturbative definition of ref.~\cite{Johnson:2019eik} (and its extension to this paper) is such that the piece of $u_0(x)$ in the $x>0$ regime is replaced by $u_0=E_0$, removing the multi-valuedness, and the string equation (\ref{eq:13}), has a nice (unique) solution. It is interesting to speculate if there are alternative non-perturbative definitions that somehow repair the multi-valuedness in the~$x<0$ regime when it occurs, maintain perturbation theory, and yield sensible solutions to some new differential equation.

\section{Final Remarks}
\label{sec:5}

In this work we have investigated and resolved some puzzles regarding the double scaled matrix models shown in refs.~\cite{Maxfield:2020ale,Witten:2020wvy} to describe certain deformations of JT gravity. In doing so, we have supplied a non-perturbative completion of the  double scaled matrix model physics that extends the proposal for ordinary JT given in ref.~\cite{Johnson:2020exp}. Studying the deformations in detail, we have uncovered an interesting phase structure, in some special cases in qualitative agreement with the semi-classical analysis of the two-dimensional dilaton gravity. Let us mention a few open questions.

\paragraph{Beyond sharp defects:} The matching between deformations of JT gravity and double scaled matrix models in refs.~\mcite{Maxfield:2020ale,Witten:2020wvy} was performed for potentials $U(\phi)$ in (\ref{eq:49}) with $\alpha_i\in(0,1/2)$. This corresponds to inserting sharp defects to the JT path integral. The issue with extending this range further is due to the fact that the decomposition of the hyperbolic surfaces in the topological expansion as done in ref.~\cite{Saad:2019lba} breaks down for $\alpha_i\in[1/2,1)$. This technical difficulty has been very recently overcome in ref. \cite{Turi}.\footnote{We thank the authors for sharing a manuscript of their work before it was submitted to arXiv.} Building on earlier studies of deformations of minimal models \cite{Moore:1991ir,Belavin:2008kv}, the authors were able to compute the leading genus string equation for $\alpha_i\in(0,1)$. For a single defect (\textit{i.e.} $r=1$ in (\ref{eq:49})) it can be written as \cite{Turi}
\begin{equation}\label{eq:110}
\mathcal{R}_0[u_0,x]=
\frac{\sqrt{u_0}}{2\pi}I_1(2\pi\sqrt{u_0})+
\sum_{n=1}^{\left \lfloor{\frac{1}{1-\alpha}}\right \rfloor}
\frac{\lambda^n}{n!}
\left[
\frac{2\pi(1-n(1-\alpha))}{\sqrt{u_0}}
\right]^{n-1}
I_{n-1}\big(2\pi(1-n(1-\alpha))\sqrt{u_0}\big)+x\ .
\end{equation}
While for $\alpha\in(0,1/2]$ there is a single term in the summation and we recover (\ref{eq:45}) as given in refs.~\mcite{Maxfield:2020ale,Witten:2020wvy}, this equation provides a very interesting and non-trivial generalization in the range $\alpha\in(1/2,1)$. 

This string equation supplies the necessary ingredients to extend our matrix model analysis from section \ref{sec:4} to $\alpha\in(0,1)$. In particular, it raises the following question: Are the non-positive spectral density~$\rho_0(E)$ and multi-valued potential $u_0(x)$ observed in figure \ref{fig:16} still present in the regime~${\alpha\in(1/2,1)}$? A simple analysis shows the answer is affirmative. For instance, taking~${\alpha=3/4}$ the threshold energy~$E_0$ obtained from the largest solution to~$\mathcal{R}_0[E_0,0]=0$ above, exhibits a zeroth order phase transition around~$\lambda\sim 0.17$. In a completely analogous way as in figure \ref{fig:16}, the potential~$u_0(x)$ becomes multi-valued in the region~$x<0$ and~$\rho_0(E)$ non-positive. Overall, this suggests that while the disc string equation~(\ref{eq:110}) becomes quite complicated in the extended range of $\alpha$, our analysis in section~\ref{sec:4} does not change that much. That being said, studying the  equation~(\ref{eq:110}) in further detail is an interesting question we would like to explore in future work.

\paragraph{Normalization of the dilaton potential:} As explained in ref. \cite{Witten:2020wvy}, the computation of the Euclidean partition function from the classical definition of the action in (\ref{eq:49}) entails some ambiguities. These have to do with different choices of operator orderings and renormalization procedures in the quantum theory. This results in an ambiguity regarding the normalization of the potential~$U(\phi)$, already seen by comparing equations (1.6) and (D.2) in refs. \cite{Witten:2020wvy} and \cite{Maxfield:2020ale} respectively. In this work we have used the normalization of ref. \cite{Witten:2020wvy}. The choice of normalization turns out being important, specially for cases in which $U(\phi)$ in (\ref{eq:49}) contains several terms, as it affects the behavior of~${U(\phi=0)}$. In particular, it determines whether the phase transition at $T=0$ discussed around~(\ref{eq:51}) is present or not in the semi-classical analysis. It would be interesting to better understand the correct normalization of the dilaton potential (see section 5.1 in ref. \cite{Turi} for progress in this direction).

\bigskip
\leftline{\bf Acknowledgments}
\noindent 
This work is  supported by the DOE grant DE-SC0011687 (USC) and NSF grant PHY-1748958 (KITP). CVJ and FR thank Henry Maxfield, Krzysztof Pilch, Joaquin Turiaci, Mykhaylo Usatyuk, Wayne Weng and Edward Witten for comments and conversations. CVJ thanks Amelia for her support.

\appendix

\section{Inverting the Abel Transform}
\label{sec:abel-inverse}

In this appendix we invert the Abel transform and derive the simple formula for~$\mathcal{R}_0[u_0,x]$ given in~(\ref{eq:48}). Our starting point is (\ref{eq:47}) with $E_0=0$
\begin{equation}
\rho_0(E)=
\frac{1}{2\pi \hbar}
\int_{0}^{E}\frac{du_0}{\sqrt{E-u_0}}
(\partial_{u_0}\mathcal{R}_0)=
\frac{(g\ast \mathcal{R}_0')(E)}{2\pi \hbar}\ , 
\end{equation}
where in the second equality we have defined $g(y)=1/\sqrt{y}$, written the integral as a convolution and the prime is a derivative with respect to $u_0$. Applying the Laplace transform, defined in the usual way
\begin{equation}
\mathcal{L}\left\lbrace f \right\rbrace (\beta)\equiv
\int_0^{+\infty}dE\,f(E)\,e^{-\beta E}\ ,
\end{equation}
and using the convolution theorem we find
\begin{equation}
\mathcal{L}\left\lbrace \rho_0 \right\rbrace (\beta)=
\frac{\mathcal{L}\left\lbrace g \right\rbrace (\beta)}{2\pi \hbar}
\mathcal{L}\left\lbrace \mathcal{R}_0' \right\rbrace (\beta)=
\frac{1}{2\hbar\sqrt{\pi \beta}}
\left[
\beta\,
\mathcal{L}\left\lbrace \mathcal{R}_0 \right\rbrace (\beta)
-\mathcal{R}_0\big|_{u_0=0}
\right]\ ,
\end{equation}
where in the second equality we used $\mathcal{L}\lbrace f' \rbrace(\beta)=\beta\,\mathcal{L}\lbrace f \rbrace(\beta)-f(0)$ and solved the Laplace transform of $g$. From the leading genus string equation (\ref{eq:46}) we can evaluate the boundary term $\mathcal{R}_0\big|_{u_0=0}=x$, given that $t_0$ must vanish so that~${u_0(0)=E_0=0}$. Solving for the Laplace transform of $\mathcal{R}_0$ we find
\begin{equation}
\mathcal{L}\left\lbrace \mathcal{R}_0 \right\rbrace (\beta)=
2\hbar\sqrt{\pi/\beta }\,
\mathcal{L}\left\lbrace \rho_0 \right\rbrace (\beta)+
x/\beta=
2\hbar\,
\mathcal{L}
\lbrace
(g\ast \rho_0)
\rbrace(\beta)+
x \mathcal{L}\lbrace 1 \rbrace(\beta)
\ ,
\end{equation}
where we have again used the convolution theorem and used $\mathcal{L}\lbrace 1 \rbrace(\beta)=1/\beta$. Applying the inverse Laplace transform we obtain the explicit formula for $\mathcal{R}_0$ in terms of $\rho_0(E)$ given in (\ref{eq:48}).

\section{Further Details of the Non-Perturbative Definition}
\label{sec:alt-derivation}

Based on the work in refs.~\cite{Dalley:1991qg,Dalley:1991xx}, in this appendix we motivate in a simple way the non-perturbative differential equation for $u(x)$ in (\ref{eq:13}) using the assumptions in (\ref{eq:60}). Differentiating the scaling relation (\ref{eq:60}) with respect to~$s$, setting~$s=1$ and using the KdV flow equation we find
\begin{equation}\label{eq:111}
u+\frac{1}{2}u'x+
\sum_{k=0}^{\infty}t_k
\frac{(2k+1)}{2(k+1)}
\widetilde{R}_{k+1}'[u]
=E_0\frac{\partial u}{\partial E_0}
\qquad \Longrightarrow \qquad
u\mathcal{R}'+\frac{1}{2}u'\mathcal{R}-\frac{\hbar^2}{4}\mathcal{R}'''=E_0\frac{\partial u}{\partial E_0}\ ,
\end{equation}
where a prime is a derivative with respect to $x$. In the second step we have used the recursion relation satisfied by $\widetilde{R}_k[u]$ (\ref{eq:59}) and we have written everything in terms of $\mathcal{R}$ defined in (\ref{eq:77}). 

Our additional condition is that whatever equation we obtain must perturbatively agree with the Hermitian matrix model, {\it i.e.,} it must have the solution ${\cal R}=0$ in $x<0$ perturbation theory. Given that $E_0$ scales the same way as $u$, one of the simplest options is that  ${\partial u}/{\partial E_0}$ is proportional to~${\cal R}'$.  Powers of ${\cal R}$ and other  powers of ${\cal R}'$ are in principle possible, which would need to be combined with powers of $x$ to achieve the correct scaling.  
However, requiring that the boundary condition for $u(x)$ at large positive~$x$ must be fixed as $u(x\rightarrow +\infty)=E_0$ (the perturbative continuity discussed in equation~(\ref{eq:19})), means that  ${\partial u}/{\partial E_0}$ must go to unity as $x\to+\infty$. Indeed, from the boundary condition $u(x\rightarrow +\infty)=E_0$ we have $\lim_{x\rightarrow +\infty} \mathcal{R}'=u'(x)\sum_{k=0}^{\infty}t_kku(x)^{k-1}+1=1$, where we used $\partial_x^n u(x)$ vanishes for $n>0$ and large positive $x$. No other combinations of  ${\cal R}$ and  ${\cal R}'$ (combined with powers of $x$, for scaling) can satisfy this additional constraint. Therefore,\footnote{While there are other more complicated scale invariant terms that could be written down that vanish to all orders in perturbation theory (such as $e^{-\hbar^2/\mathcal{R}}$) , it is not clear if any of them can be produced from a matrix model, as is the case for the one given in equation (\ref{eq:112}).} 
\begin{equation}\label{eq:112}
\frac{\partial u}{\partial E_0}=
\mathcal{R}'\ .
\end{equation}
Using this we can multiply both sides in (\ref{eq:111}) by $\mathcal{R}$, giving a total derivative, then integrate and obtain:
\begin{equation}\label{eq:113}
(u-E_0)\mathcal{R}^2+\frac{\hbar^2}{2}\mathcal{R}\mathcal{R}''+\frac{\hbar^2}{4}(\mathcal{R}')^2=0\ ,
\end{equation}
where we have fixed the integration constant to zero, so that $\mathcal{R}=0$ (the Hermitian model string equation) is a solution. We have arrived at the non-perturbative string equation (\ref{eq:13}) for $u(x)$. 

Some interesting features of the physics are apparent from this derivation (some of which have been studied in ref.~\cite{Johnson:1992wr}). For example, equation~(\ref{eq:112}) shows that as $E_0$ changes (as it will under some deformations) by some amount $\delta E_0$, the corresponding change to the function in the asymptotic region (because ${\cal R}'\to1$ there) is  $\delta u=\delta E_0$, preserving the boundary condition as it should. More generally, equation~(\ref{eq:112}) is a new non-perturbative flow equation (joining the KdV ones) for $u$ as a function of $E_0$. Recall also that  the parameter $E_0$ (the end of the matrix model spectrum) defines ``the wall''. It arises from the potential that defines the probability measure, now  as a function of $(MM^\dagger+E_0)\ge E_0$. In the limit of $E_0\rightarrow -\infty$ the model is increasingly equivalent to an Hermitian matrix model (arbitrarily negative eigenvalues are now allowed),  and indeed the dominant part of the equation (\ref{eq:113}) in that limit  is simply $E_0{\cal R}^2=0$, which  forces $u(x)$ to solve the Hermitian matrix model equation~$\mathcal{R}=0$. 

\section{Deriving a B\"{a}cklund Transformation}
\label{sec:Backlund}

In this appendix we derive a B\"{a}cklund transformation that relates solutions to the differential equation (\ref{eq:13}) when the right-hand side has the constant $\hbar^2\Gamma^2$ instead of zero. This is a slight generalization of the derivation in refs.~\cite{Dalley:1992br,Carlisle:2005mk}, where this transformation was derived for $E_0=0$.

Let us start by assuming we have $u(x)$, a solution to the differential equation (\ref{eq:13}). We can use this solution to define two functions $v_\pm(x)$ according to
\begin{equation}\label{eq:4}
v_\pm=\frac{\hbar}{2}
\left(\frac{ \mathcal{R}'[u,x]\mp 2 \Gamma}
{\mathcal{R}[u,x]}\right)
\qquad \Longrightarrow \qquad
X_\pm\equiv \frac{\hbar}{2}\mathcal{R}'[u,x]
-v_\pm	 \mathcal{R}[u,x]
\mp \hbar \Gamma
=0 \ .
\end{equation}
Using $X_\pm=0$ together with the fact $u(x)$ satisfies the differential equation (\ref{eq:13}) (with $\hbar^2\Gamma^2$ on the right-hand side) we have the following trivial identity
\begin{equation}
0=X_\pm\left(
X_\pm \pm  2\hbar \Gamma
\right)-\hbar \mathcal{R}X_\pm'=
(u-E_0)\mathcal{R}^2
-\frac{\hbar^2}{2}\mathcal{R}\mathcal{R}''
+\frac{\hbar^2}{4}
(\mathcal{R}')^2
-\hbar^2\Gamma^2\ ,
\end{equation}
that we can use to solve for $u(x)$ in terms of $v_\pm(x)$
\begin{equation}\label{eq:3}
u(x)=
v_{\pm}^2(x)+\hbar v_{\pm}'(x)+E_0\ .
\end{equation}
This is the Miura transformation, mapping between the KdV and mKdV hierarchies, generalized to non-zero~$E_0$. The differential equations satisfied by the functions~$v_\pm(x)$ can be obtained combining~(\ref{eq:3}) and~(\ref{eq:4}), so that we find
\begin{equation}\label{eq:5}
\sum_{k=1}^{\infty}t_k
\mathcal{S}_k[v_\pm]
-v_{\pm}(x+t_0)=\hbar(\pm \Gamma-1/2)
\ .
\end{equation}
where we have defined
\begin{equation}\label{eq:9}
\mathcal{S}_k[v_\pm]\equiv 
\frac{\hbar}{2}
\widetilde{R}'_k[v_{\pm}^2+\hbar v_{\pm}'+E_0]
-
v_{\pm}
\widetilde{R}_k[v_{\pm}^2+\hbar v_{\pm}'+E_0]\ .
\end{equation}
This is the string equation in the mKdV hierarchy (arising from unitary \cite{Periwal:1990qb} and Hermitian multi-cut~\cite{Crnkovic:1990mr} matrix models) generalized to non-zero $E_0$.

Let us introduce an additional notation, by adding a subscript to $u(x)$ and $v_\pm(x)$ that indicates the parameter that appears on the right hand side of their respective differential equations. The solutions to (\ref{eq:13}) and (\ref{eq:5}) in this notation are given by
\begin{equation}\label{eq:8}
u(x)\equiv u_\Gamma(x)\ ,
\qquad \qquad
v_\pm(x) \equiv v_{\pm\Gamma-1/2}(x)\ .
\end{equation}
We now note the functional $\mathcal{S}_k[v(x)]$ in (\ref{eq:9}) is odd under reflections, \textit{i.e.} $\mathcal{S}_k[-v(x)]=-\mathcal{S}_k[v(x)]$. For instance, if we consider $k=1$ so that $\widetilde{R}_{k=1}[f(x)]=f(x)$ we have
\begin{equation}
\mathcal{S}_{k=1}[v(x)]=
\frac{\hbar^2}{2}v''(x)-E_0 v(x)-v(x)^3=
-\mathcal{S}_{k=1}[-v(x)]\ .
\end{equation}
While this property was noted in refs.~\cite{Periwal:1990qb,Dalley:1992br} for $E_0=0$ it is non-trivially extended to non-zero $E_0$. For any arbitrary value of $k>1$ it is straightforward to check it continues to hold. This turns out to be very useful, as it gives us another way of relating solutions to the differential equation in~(\ref{eq:5}). Given any solution $v_c(x)$ (with constant $\hbar c$ on the right hand side), we can use the reflection property satisfied by $\mathcal{S}_k$ to generate a solution with opposite constant, \textit{i.e.} $v_{-c}(x)=-v_c(x)$.

We now have all the ingredients necessary to derive the B\"{a}cklund transformation for $u_\Gamma(x)$, shown below in (\ref{eq:24}). Let us prove it for $(\Gamma-1)$, the other case being completely analogous. The starting point is (\ref{eq:3}), which in the notation in (\ref{eq:8}) becomes
\begin{equation}\label{eq:10}
u_\Gamma(x)=
v^2_{\pm \Gamma-1/2}(x)
+\hbar v'_{\pm \Gamma-1/2}(x)
+E_0\ .
\end{equation}
Changing $\Gamma\rightarrow \Gamma-1$ and using $v_{-c}(x)=-v_c(x)$ we can transform this to
\begin{equation}
u_{\Gamma-1}(x)=
v^2_{\Gamma-1/2}(x)
-\hbar v'_{\Gamma-1/2}(x)
+E_0\ ,
\end{equation}
which combined with (\ref{eq:10}) with the upper sign gives
\begin{equation}
u_{\Gamma-1}(x)=
2v^2_{\Gamma-1/2}(x)
-u_\Gamma(x)+2E_0\ .
\end{equation}
The B\"{a}cklund transformation is obtained after rewriting the first term in this expression using (\ref{eq:4}):
\begin{equation}\label{eq:24}
u_{\Gamma \pm 1}(x)=
\frac{\hbar^2}{2}\left(
\frac{ \mathcal{R}'[u_\Gamma,x]\pm 2 \Gamma}
{\mathcal{R}[u_\Gamma,x]}
\right)^2-u_\Gamma(x)
+2E_0 \ .
\end{equation}

This is a useful transformation in a number of situations. For example, sometimes it is easier to solve the string equation at another value of $\Gamma$ (because for example, it has a shallower potential well), and then afterwards apply the transformation to change $\Gamma$. This was sometimes done in ref.~\cite{Johnson:2020exp} as a convenience ({\it e.g.,} solving for $\Gamma=1$ and then reducing to $\Gamma=0$ afterwards), although the current paper did not need to employ this method since the numerical methods used have since become more powerful for solving directly at $\Gamma=0$. It is also possible to imagine using this transformation successively to reduce $\Gamma$  from some high integer value down to zero, having started with it large, or even formally infinite. 

Finally, note that in the 't Hooft limit of section~\ref{sub:4.3} where the solution is algebraic,  it is easy to see that there is no path to smaller $\Gamma$ using these transformations. Since here $u_\Gamma(x)$ is $\hat{u}(x)$, a solution to equation~(\ref{eq:22}) valid in the assumed limit, the same limit should be used in the transformation. This means that the transformation becomes:
\begin{equation}\label{eq:24a}
u_{\Gamma \pm 1}(x)=
\frac{2q^2}
{\mathcal{R}^2_0[\hat{u},x]}
-\hat{u}(x)
+2E_0  =\hat{u}(x)\ ,
\end{equation}
where in the last step equation~(\ref{eq:22}) was used to write $q^2/{\cal R}^2_0[\hat{u},x] = \hat{u}-E_0$. In other words, the~'t~Hooft limit solution $\hat{u}(x)$ is  invariant under the B\"acklund transformation.

\bibliography{sample,cvj_JT_gravity}
\bibliographystyle{JHEP}

\end{document}